\documentclass[twocolumn]{aastex631}

\usepackage{color}
\usepackage{amsmath}
\usepackage{natbib}

\hypersetup{breaklinks=true, colorlinks=true, urlcolor={blue}, citecolor={blue}, linkcolor={blue}}

\newcommand{\nustar}{\emph{NuSTAR}}

\newcommand{\nicer}{\emph{NICER}}

\def \src {SGR~1935$+$2154}

\begin{document}
\title{Rapid Spectral Evolution of SGR 1935+2154 During its 2022 Outburst}

\author[0000-0001-8551-2002]{Chin-Ping Hu}
\affiliation{Department of Physics, National Changhua University of Education, Changhua, 50007, Taiwan}
\email{cphu0821@gm.ncue.edu.tw}
\author[0000-0002-9249-0515]{Zorawar Wadiasingh}
\affiliation{Department of Astronomy, University of Maryland, College Park, Maryland 20742, USA}
\affiliation{Astrophysics Science Division, NASA Goddard Space Flight Center,Greenbelt, MD 20771, USA}
\affiliation{Center for Research and Exploration in Space Science and Technology, NASA/GSFC, Greenbelt, Maryland 20771, USA}
\author[0000-0002-6089-6836]{Wynn~C.~G.~Ho}
\affiliation{Department of Physics and Astronomy, Haverford College, 370 Lancaster Avenue, Haverford, PA 19041, USA}
\author[0000-0003-4433-1365]{Matthew~G.~Baring}
\affiliation{Department of Physics and Astronomy - MS 108, Rice University, 6100 Main Street, Houston, Texas 77251-1892, USA}
\author[0000-0002-7991-028X]{George~A.~Younes}
\affiliation{Astrophysics Science Division, NASA Goddard Space Flight Center, Greenbelt, MD 20771, USA}
\affiliation{Center for Space Sciences and Technology, University of Maryland, Baltimore County, Baltimore, MD 21250, USA}
\author[0000-0003-1244-3100]{Teruaki Enoto}
\affiliation{Department of Physics, Kyoto University, Kitashirakawa Oiwake, Sakyo, Kyoto 606-8502, Japan}
\affiliation{RIKEN Center for Advanced Photonics (RAP), 2-1 Hirosawa, Wako, Saitama 351-0198, Japan}
\author[0000-0002-6449-106X]{Sebastien Guillot}
\affiliation{Institut de Recherche en Astrophysique et Plan\'{e}tologie, UPS-OMP, CNRS, CNES, 9 avenue du Colonel Roche, BP 44346, Toulouse Cedex 4, 31028, France}
\author[0000-0002-3531-9842]{Tolga G\"uver}
\affiliation{Istanbul University, Science Faculty, Department of Astronomy and Space Sciences, Beyaz\i t, 34119, Istanbul, Turkey}
\affiliation{Istanbul University Observatory Research and Application Center, Istanbul University 34119, Istanbul Turkey}
\author[0009-0006-3567-981X]{Marlon L. Bause}
\affiliation{Max Planck Institut für Radioastronomie, Auf dem Hügel 69, 53121 Bonn, Germany}
\author[0000-0002-0254-5915]{Rachael~Stewart}
\affiliation{Department of Physics, The George Washington University, Washington, DC 20052, USA}
\author[0000-0002-3905-4853]{Alex Van Kooten}
\affiliation{Department of Physics, The George Washington University, Washington, DC 20052, USA}
\author{Chryssa~Kouveliotou}
\affiliation{Department of Physics, The George Washington University, Washington, DC 20052, USA}

\correspondingauthor{C.-P. Hu}

\begin{abstract}
During the 2022 outburst of \src, a Fast-Radio-Burst-like event (FRB 20221014A) and X-ray activities occurred between two spin-up glitches, suggesting these glitches may connect to multiwavelength phenomenology. 
However, the mechanisms altering the magnetar's magnetosphere to enable radio emission remain unclear. 
This study presents high-cadence \nicer\ and \nustar\ observations revealing spectral changes in burst and persistent emission.
Hardness ratio and spectral analysis reveal significant changes during an ``intermediate flare'' 2.5 hours before FRB 20221014A.
This 40-s flare, releasing $>(6.3\pm0.2)\times10^{40}$ erg, coincides with a rapid spectral softening in both burst and persistent emission and a notable decrease in burst occurrence rate. 
The intermediate flare is bright enough to be detected if placed at a few Mpc, and would appear as a fast X-ray transient. 
This implies that the connection between magnetar X-ray activity and FRBs can be observed in the local Universe.
Post-flare burst spectra peak near 5 keV, resembling the characteristics of the FRB-associated X-ray burst of 2020.
Such change persisted for a few hours, implying magnetospheric evolution on similar timescales.
However, no radio emission was detected from post-flare bursts, suggesting that FRB emission requires conditions beyond peculiar short bursts. 
The burst waiting times exhibit a broken power-law distribution, likely resulting from contamination by enhanced persistent emission.
Although the bursts appear randomly distributed in the spin phase, the hardness ratio profile as a function of spin phase follows that of the persistent emission, indicating that X-ray bursts originate at low altitudes. 
\end{abstract}
\keywords{High energy astrophysics (739); Compact objects (288); Neutron Stars (1108); Magnetars (992); Radio transient sources (2008)}

\submitjournal{ApJ}
\received{2025 March 06}
\accepted{2025 June 27}

\section{Introduction}\label{introduction}
Fast Radio Bursts (FRBs), one of the most puzzling astrophysical phenomena in this decade, represent a class of extragalactic radio pulses distinguished by their high luminosity, short duration  (lasting less than a second), and broadband, coherent radio emission \citep{Petroff2019, PetroffHL2022, Bailes2022, Zhang2023}. 
The first likely genuine FRB, known as the Lorimer burst, was detected in 2001 with the Parkes Radio Telescope \citep[][but see \cite{Linscott1980}]{LorimerBM2007}. 
After the Canadian Hydrogen Intensity Mapping Experiment (CHIME) began its FRB program in 2018, the number of detected FRBs increased rapidly \citep{CHIME_Catalog_2021}.
FRBs are conventionally classified into two classes: repeaters and apparent non-repeaters \citep{Spitler2016}. 
Despite numerous theoretical models proposed to explain these transient radio events, their progenitors remain unidentified \citep{Platts2019,Zhang2023}.

Magnetars, a subclass of neutron stars, are characterized by their extremely strong magnetic fields and diverse X-ray phenomena \citep{DuncanT1992,Paczynski1992}. 
These phenomena include sporadic short X-ray bursts ranging from milliseconds to several seconds, long-term outbursts with flux enhancements over months to years, and a variable spin-down rate in their rotation periods \citep{KaspiB2017,EnotoKS2019}.
During an outburst, a magnetar exhibits enhanced activity, emitting a series of short X-ray bursts and potentially giant flares, the latter releasing total energies on the order of $10^{44}$ erg within several minutes \citep[see, e.g.,][]{MazetsGG1979,EvansKL1980, HurleyCM1999, HurleyBS2005, BurnsSH2021, BeniaminiWT2024}.

\begin{table*}[t]
 \caption{Observations used in this analysis}
 \label{tab:obs_log}
 \begin{tabular}{lcccc}
 \hline
 Observatory & ObsID & Start & End & Exposure (ks)\\
 \hline
\nustar\ & 80802317002 & 2022-10-14 02:01:09 & 2022-10-15 05:06:09 & 49 \\
\nustar\ & 80802317004 & 2022-10-26 08:26:09 & 2022-10-27 10:21:09 & 50 \\
\hline
\nicer\ & 5576010102 & 2022-10-14 03:34:37 & 2022-10-14 22:20:40 & 10.3\\
\nicer\ & 5576010103 & 2022-10-14 23:58:20 & 2022-10-15 22:02:40 & 7.8 \\
\hline
 \end{tabular}
\end{table*}

Soon after the FRB discovery, neutron stars and magnetars emerged as one of the most promising candidates for their origin \citep[e.g.,][]{PopovP2010, KeaneSK2012, Lyubarsky2014, MuraseKM2016, Lyutikov2017, Beloborodov2017, Kumar2017, Wadiasingh2019,Wadiasingh2020, Zhang2020}. 
The connection between magnetar giant flares and FRBs was proposed by \citet{PopovP2010}, and many models have been developed since then.
However, in these models, the exact location of the radio emission or trigger remains uncertain. 
FRBs have been suggested to be generated by magnetic reconnection in the neutron star magnetosphere \citep{Lyutikov2020, Lyutikov2021}, curvature radiation near the neutron star surface \citep{Kumar2017}, or synchrotron maser emission at the termination shock generated by a giant flare \citep{Lyubarsky2014, Metzger2019}.

In 2020, the first direct evidence connecting magnetars to FRBs was found as the detection of FRB-like emission (FRB 20200428) from an outbursting galactic magnetar \src\ observed by CHIME and STARE2 \citep{CHIME2020_SGR1935, BochenekRB2020}.
This FRB exhibited a double-peaked feature with radio fluences of 420 kJy~ms and 220 kJy~ms.
\src\ is a magnetar with a rotation period of $P = 3.24$ s and a long-term period derivative of $\dot{P} = 1.43 \times 10^{-11}$ s s$^{-1}$ \citep{Israel2016, Younes2015}, implying a surface magnetic field strength of $B = 2.2 \times 10^{14}$ G and a characteristic age of $\tau = 3.6$ kyr. 
It is one of the most active magnetars in recent years.
The FRB-associated X-ray burst, characterized by hard X-ray non-thermal emission, was observed by the INTEGRAL, Konus-Wind, and HXMT missions \citep{MereghettiSF2020, RidnaiaSF2021, LiLX2021} and was not a giant flare as some of the above models predicted.

Among these bursts, 24 of them were simultaneously detected by the Neutron Star Interior Composition Explorer (\nicer) and the Fermi Gamma-ray Burst Monitor (GBM).
These bursts exhibited spectral behaviors strikingly different from that of the FRB-associated X-ray burst, suggesting that the FRB-associated burst originated from a unique or uncommon location \citep{YounesBK2021}, and implying special conditions are necessary for FRB generation or escape.

Later, in October 2020, \src\ emitted several \textcolor{blue}{bright} radio bursts over a few days \citep{Kirsten2020}, followed by pulsed radio emission lasting approximately a month \citep{ZhuXZ2023}. 
Interestingly, no X-ray bursts or outbursts were detected around these events. 
Instead, a potential spin-down glitch was observed although this glitch might have been a combination of a spin-up glitch and rapid spin-down \citep{Younes2023, HuNE2024}. 
The fast spin down may have generated a wind that combed out the magnetic field lines, temporarily changing the magnetospheric field geometry, and enhancing the probability of FRBs from short bursts.

The most recent outburst of \src\ was alerted by a short X-ray burst detected at Coordinated Universal Time (UTC) 21:28:25.8 on October 10 \textcolor{blue}{in 2022} with the International Gamma-Ray Astrophysics Laboratory \citep{MereghettiGF2022}. 
A few days later, on October 14, 2022, at 19:21:47 UTC (topocentric time), an FRB-like event (FRB 20221014A) with multiple radio peaks was detected by CHIME and the Robert C.~Byrd Green Bank Telescope (GBT, \citealt{Dong_CHIME_2022, MaanLS2022}). 
High-cadence joint observations by \nicer\ and  Nuclear Spectroscopic Telescope Array (\nustar) revealed a short-term X-ray outburst that occurred two hours before the FRB, with a decay timescale of around one day, along with a burst forest that took place soon after the rising edge. 
Interestingly, after removing the burst emission, the pulsation of \src\ showed two spin-up glitch events, with the first occurring approximately 4.5 hours before FRB 20221014A and the second around 4.5 hours after the event  \citep{HuNE2024}. 
This suggests that rotation and superfluid dynamics could partially contribute to the multi-wavelength activity of magnetars and FRBs.
The post-glitch spin-down rate remains approximately four times higher than the long-term value, and the spectra have been confirmed to only have slight variability \citep{IbrahimBC2024}.
The LIGO-Virgo-KAGRA collaborations have searched for possible gravitational wave signals from these glitches and estimated a 50\% upper limit of energy release of approximately $10^{50}$~erg \citep{LVK2024_SGR1935}.

As a follow-up research of \citet{HuNE2024}, we focus on the timing and spectral properties of short X-ray bursts and persistent emission detected during the activity in 2022. 
Section \ref{observation} describes the observational data collected for this study, including data from \nicer\ and \nustar.
The detailed analysis results, including the evolution of short X-ray bursts and the persistent emission, burst spectroscopy, and burst statistics, are presented in Section \ref{analysis}.
The possible origin of the changes in the hardness ratio, the interpretation of the burst distribution, and the origin of the FRB emission are discussed in Section \ref{discussion}.
Finally, we summarize our work in section \ref{summary}.

\begin{figure*}[h!]
\includegraphics[width=0.95\textwidth]{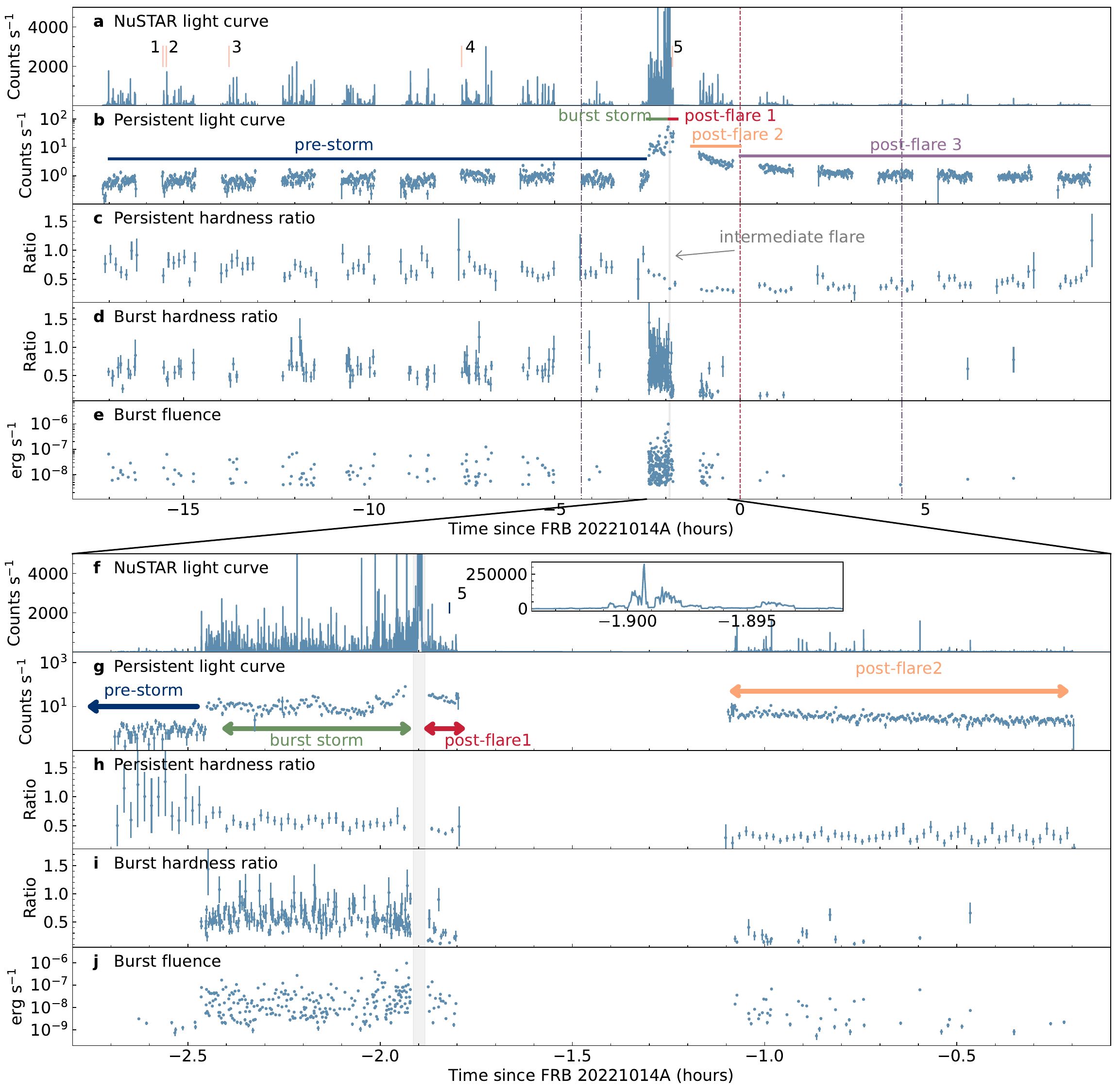}
\caption{Light curve, hardness ratio, and burst fluence observed with \nustar. Panel \textbf{a} displays the \nustar\ light curve with a time bin size of 0.01 s, while panel \textbf{b} presents the persistent light curve with a time bin size of 64 s. In panel \textbf{a}, the vertical pink lines with numbers 1 to 5 highlight five luminous bursts detected simultaneously by \nicer\ and \nustar, with burst 5 being primarily detected in \nustar's mode 06 data. Panel \textbf{c} reveals the variability of the hardness ratio for the persistent emission, using a time bin size of 512 s, and panel \textbf{d} shows the hardness ratio for individual bursts with a fluence greater than $5\times 10^{-9}$ erg. Panel \textbf{e} shows the fluence of all detected bursts. A vertical red-dashed line marks the onset of FRB 20221014A, while the purple dash-dotted lines indicate the timings of two glitches. Panels \textbf{f} to \textbf{j} provide zoomed-in views of panels \textbf{a} to \textbf{e}, respectively, with bin sizes for the persistent light curve in panel \textbf{g} and the hardness ratio in panel \textbf{h} adjusted to 16 s and 64s, respectively. The time interval of the intermediate flare is shaded in gray. Periodic gaps in data are due to \nustar\ occultation by the Earth.
}\label{fig:burst_evolution_all}
\end{figure*}

\section{Observations and Data Reduction}\label{observation}
\subsection{\nicer\ Observation}
\nicer\ is a non-imaging X-ray observatory aboard the International Space Station that has a large effective area of 1900 cm$^2$ at 1.5 keV, a high time resolution of $<300$~ns \citep{GendreauAA2016}.
Among 56 co-aligned X-ray optics to focus X-rays onto silicon-drift detectors, 52 are currently in operation.

We initiated a sequence of \nicer\ observations starting at 17:32:40 UTC on October 12.
Two observations were jointly observed within the \nustar\ time coverage and used in this analysis (see Table \ref{tab:obs_log}).
We reprocessed the \nicer\ data using NICERDAS version 9, part of the HEASOFT version 6.30.1 package, and the calibration database version 20210720. 
The raw data was calibrated and screened via the \texttt{nicerl2} pipeline under standard parameters, yielding cleaned event lists. 
We then corrected the photon arrival times to the solar system's barycenter using the \texttt{barycorr} tool with the JPL ephemeris DE405.

For the timing analysis, photons below 2 keV were excluded due to significant absorption, and those above 8 keV were omitted due to \nicer's reduced sensitivity at high energies.
We extracted the burst light curves using the \texttt{nicerl3-lc} command and the \nicer\ spectra using \texttt{nicerl3-spect}. 
For joint spectral fitting with \nustar, we employed the SCORPEON model, setting the background model type to \texttt{file}.  
We utilized photons in the energy range of 1 to 10 keV for spectral analysis.

\subsection{\nustar\ Observation}
\nustar\ is an imaging hard X-ray telescope designed to observe in the 3 -- 79 keV energy range.
It consists of two co-aligned focal plane modules, FPMA and FPMB, which use conical Wolter-I optics to achieve a point spread function with a half-power diameter of approximately 60\arcsec.
These modules offer a $12\arcmin \times 12\arcmin$ field of view, an energy resolution of 0.4 keV at 10 keV, and a timing resolution of 65 ms \citep{BachettiMG2021}.

We carried out two observations on \src. 
The first one began on October 14 at 02:01:09 UTC and covered from 17 hours before to 10 hours after FRB 20221014A.
The second was scheduled on October 26, roughly 12 days after the first observation.
The data were reprocessed using NUSTARDAS version 2.1.2 with the calibration database version 20211020. 
We calibrated and screened the data with the \texttt{nupipeline} script, applying a status expression of ``\texttt{statusexpr='STATUS==b0000xxx00xxxx000'}'' to address severe deadtime issues during the burst storm. 
Subsequent data products, including images, event lists, light curves, and spectra, were generated using the \texttt{nuproducts} tool.

We collected source photons from a circular region with a radius of 0.85\arcmin\  centered on the source, ensuring that approximately 85\% of the source X-ray photons were collected. 
Background spectra were extracted from a circular source-free region with a 2.5\arcmin\ radius located at R.A.~19:34:46.6029 and Decl.~+21:46:55.710.

Additionally, we observed a few bursts in the mode 06 data (SCIENCE\_SC: attitude reconstruction from the spacecraft bus star trackers), which may display multiple centroids from the same target due to imprecise aspect reconstruction. 
The photons collected during these time intervals are not included in the regular mode 01 (SCIENCE: normal observing scientific mode) pipeline output and must be processed separately.
We split the mode 06 event files into distinct ones for each combination of the Camera Head Unit (CHU), CHU1, CHU2, and CHU3, using the \texttt{nusplitsc} command. 
When examining the images, we found that \src\ appears as a point-like source in CHU1 and CHU2, where a few bursts are detectable.
Therefore, we extracted the light curves and spectra for these bursts using nuproducts.

\section{Analysis and Results}\label{analysis}

\subsection{Light Curve and Burst Property Evolution}
We performed a burst search in both \nicer\ (2--8~keV) and \nustar\ (3--79 keV) event lists using the Bayesian block technique\footnote{\href{https://docs.astropy.org/en/stable/api/astropy.stats.bayesian_blocks.html}{https://docs.astropy.org/en/stable/api/astropy.stats.bayesian\_blocks.html}}. 
This approach identifies local variability by segmenting the data into time blocks of varying sizes to detect significant changes in flux \citep{ScargleNJ2013}. 
Initially, we choose blocks lasting less than 4 s and then merged nearby blocks, as bright and extended bursts are typically recorded as several consecutive blocks.
Then, we evaluated whether the total count in each block could be due to random Poisson fluctuations, considering the count rate roughly three seconds before and after each block.
During the peak of the burst storm (around $-2.5<t<-1.9$ hours, where $t=0$ is defined as the occurrence time of FRB20221014A), flux peaks can occur in short time intervals. 
This complexity makes it challenging to determine whether we see a single burst with multiple peaks or several individual bursts. 
In such cases, we classify them as separate bursts unless the flux between the two peaks remains remains approximately 10\% higher than the surrounding persistent count rate.
A total of 633 burst candidates were detected with \nustar\ and cataloged in Appendix \ref{sec:appendix1}. 
An approximately 40-s long intermediate flare was observed in the \nustar\ data set (see Figure \ref{fig:burst_evolution_all}). 
Note that the intermediate flare itself, and several oscillation‐like features occurring within one minute before the intermediate flare, are excluded from this table.
During this entire activity, we detected 368 bursts in the \nicer\ data set \citep{HuNE2024, TsuzukiTH2024}. These events will be presented in an upcoming \nicer\ magnetar short X-ray burst catalog (Chu et al.~in prep).

\begin{table*}[ht]
 \caption{Parameters of five bursts that are simultaneously detected with \nicer\ and \nustar. The errors quoted in this paper are 1$\sigma$ uncertainties. }
 \label{tab:bursrs_parameters}
 \begin{tabular}{llccccc}
 \hline
Model & Parameters & Burst1 & Burst2 & Burst3 & Burst4 & Burst5 \\
 \hline
& Time (hour) & $-15.5658$ & $-15.4827$ & $-13.7899$ & $-7.5215$ & $-1.8362$  \\
& T90 (s)  & 0.62 & 1.97  & 1.09 & 0.62 & 2.90 \\
\hline
CPL & $N_{\textrm{H}}$ ($10^{22}$ cm$^{-1}$)  & \multicolumn{5}{c}{$(3.6\pm0.1)$} \\
 &$\Gamma$  & $-0.4\pm0.3$ & $-0.5\pm0.1$ & $-0.2\pm0.2$ & $0.0\pm0.2$ & $0.65\pm0.07$ \\
 &$E_f$ (keV)  & $6_{-1}^{+2}$ & $5.2_{-0.5}^{+0.6}$ & $5.9_{-0.9}^{+1.2}$ & $6_{-1}^{+2}$ & $3.5\pm0.2$ \\
 \hline
\multicolumn{7}{l}{Fit statistic $=1407.1$ with 1358 degrees of freedom, null hypothesis probability of 0.018} \\
\hline
2BB & $N_{\textrm{H}}$ ($10^{22}$ cm$^{-1}$) & \multicolumn{5}{c}{$(2.6\pm0.2)$} \\
 &$kT_1$ (keV) & $1.8\pm0.2$  & $2.1\pm0.2$  & $1.6\pm0.2$ & $1.5\pm0.2$ &$1.00\pm0.03$ \\
 &$R_1$ (km) & $7.3_{-0.9}^{+2.0}$ & $6.8_{-0.3}^{+0.6}$ & $6.6_{-0.6}^{+1.3}$ & $8.3_{-0.9}^{+2.0}$ & $32.9_{-0.9}^{+0.8}$\\
 &$kT_2$ (keV) & $>5.2$ & $6.8_{-1.5}^{+2.6}$ & $4.6_{-0.8}^{+1.7}$ & $4.4_{-0.8}^{+0.7}$ & $2.6_{-0.2}^{+0.3}$ \\
 & $R_2$ (km) & $<3.2$ & $0.7_{-0.3}^{+0.8}$ & $1.2_{-0.5}^{+1.2}$ & $1.4_{-0.6}^{+1.5}$ &  $3.9_{-0.8}^{+1.7}$\\
 \hline
\multicolumn{7}{l}{Fit statistic $=1388.8$ with 1353 degrees of freedom, null hypothesis probability of 0.046} \\
\hline
 \end{tabular}
\end{table*}

We derived the burst properties using the photon events for faint bursts with durations shorter than 10 ms. 
On the other hand, the burst properties of those bursts with durations longer than 10 ms obtained with \nustar\ are derived using the deadtime corrected light curves created with \texttt{nuproducts} command. 
For each burst, we determined the T90 duration using photon counts, where T90 is defined by the interval needed to accumulate 5\% to 95\% of the total fluence \citep{KouveliotouMF1993}.
The hardness ratio of a burst detected by \nustar\ is defined as the photon count ratio of the 10--79 keV to the 3--10 keV range.
Due to the limited number of photons collected in faint bursts, we cannot perform spectral analysis of all bursts. 
Therefore, we estimate the fluence of 3--79 keV using WebPIMMs\footnote{\url{https://heasarc.gsfc.nasa.gov/cgi-bin/Tools/w3pimms/w3pimms.pl}} and assume an $N_\textrm{H}=3.6\times10^{22}$~cm$^{-2}$ and a power-law spectrum with a photon index $\Gamma=1.3$, roughly the same as the $\Gamma$ value of the averaged burst emission (see Section \ref{sec:spectral_analysis}).
In this analysis, bursts detected with \nicer\ were not included due to the different energy range compared to \nustar.

Figure \ref{fig:burst_evolution_all}a displays the deadtime corrected \nustar\ light curve with 10 ms binning, alongside the light curve and hardness ratio for the persistent (non-burst) emission (Figures \ref{fig:burst_evolution_all}b and \ref{fig:burst_evolution_all}c), as well as the evolution of the hardness ratio and fluences for the bursts (Figures \ref{fig:burst_evolution_all}d and \ref{fig:burst_evolution_all}e).
The high-resolution light curve (Figure \ref{fig:burst_evolution_all}a) suggests that \src\ was burst active since the start of this \nustar\ observation, and it reached a maximum at $t\approx -2$ hours since FRB 20221014A. 
The persistent light curve is binned with a 64-s interval, whereas the persistent hardness ratio is binned with a 512-s interval.
The persistent flux exhibited a short-term outburst, with a sharp increase at $t\approx-2.5$ hours, followed by a burst storm. 
The flux peaked at $t\approx-1.9$ hours and then gradually decreased over a timescale of approximately 10 hours.
At the same time, the persistent hardness ratio decreased after the outburst, reaching its minimum around FRB 20221014A ($t\approx0$ hours), before gradually increasing again.
The hardness ratio of the bursts shown in Figure \ref{fig:burst_evolution_all}d is calculated from those with a fluence greater than $5\times10^{-9}$ ergs s$^{-1}$, which corresponds roughly to the median fluence of all bursts, as the uncertainty in the lower-fluence bursts is large. 
The hardness ratio of the bursts exhibits a pattern similar to that of persistent emission, although individual bursts may show significant diversity.

To observe the detailed variability throughout the evolution, we present a zoom-in view of its evolution in Figures \ref{fig:burst_evolution_all}f -- \ref{fig:burst_evolution_all}j, using a 16-s bin size for the persistent light curve and hardness ratio.
A 40-s long intermediate flare, shaded in gray in Figure \ref{fig:burst_evolution_all}, was detected half hour after the rising edge of the short-term outburst, $t\approx-1.9$ hours (2.5 hours after the first glitch), during which \nustar\ encountered severe deadtime issues due to a large number of incident photons.
The deadtime-corrected light curve for the flare is inserted in Figure \ref{fig:burst_evolution_all}f. 
The 3--79 keV count rate of this flare exceeds 200,000 counts s$^{-1}$. 
However, the reliability of this measurement is limited, as \nustar\ was operational for only 0.9\% of the burst time interval due to the deadtime issue.
The variation in the hardness ratio is also observable in the zoom-in view of persistent and burst emission.
We observed a sharp decrease in the burst hardness ratio at $t\approx-1.9$ hours, coinciding with the occurrence of the intermediate flare.

\begin{figure*}[t]
    \centering
    \begin{minipage}{0.49\linewidth}
        \includegraphics[width=0.95\textwidth]{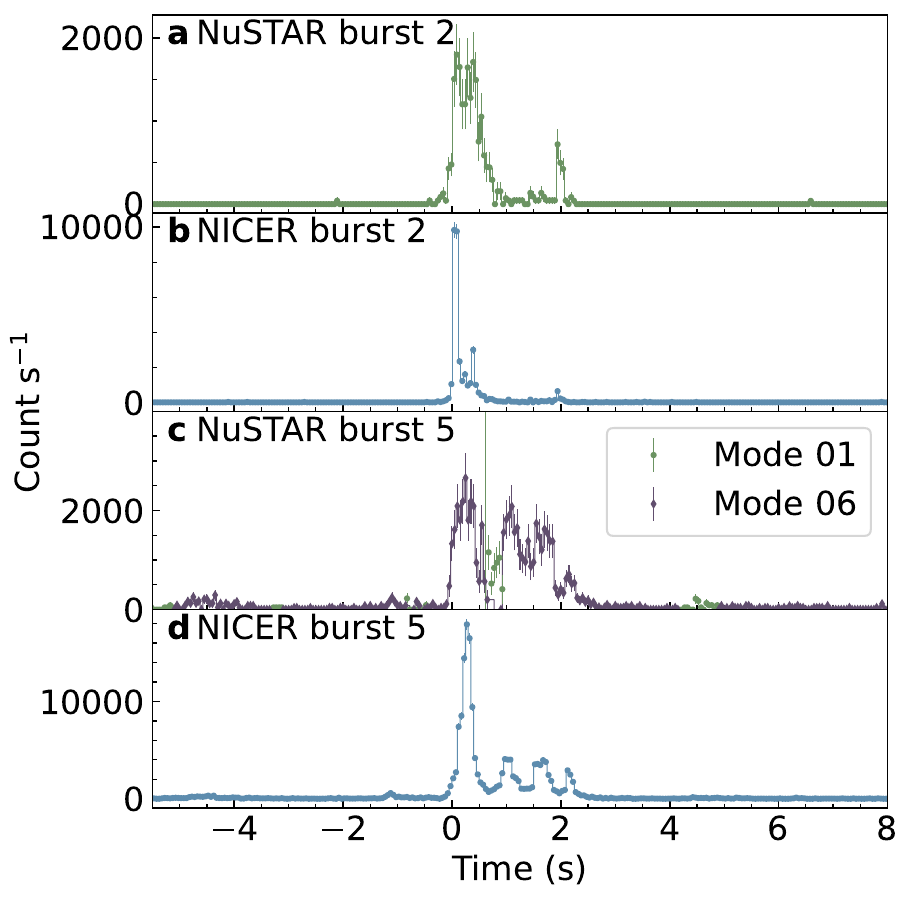}
    \end{minipage}
    \begin{minipage}{0.49\linewidth}
        \includegraphics[width=1.0\textwidth]{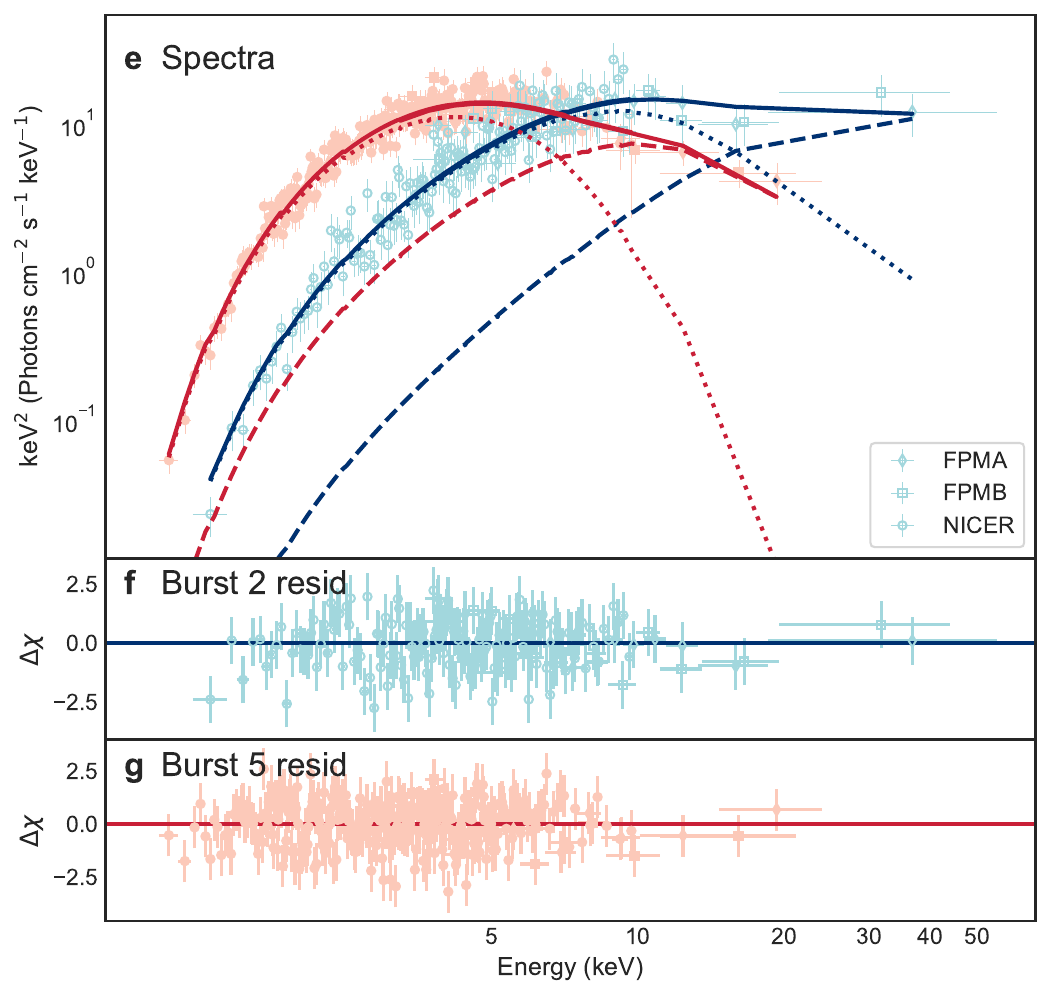}
    \end{minipage}
    \caption{Light curves and broadband spectra of bursts 2 and 5 detected by \nicer\ and \nustar. Panel \textbf{a} displays the \nustar\ 3–79 keV light curve for burst 2 with a time bin size of 0.05 s, where time zero marks the burst's fiducial start. Panel \textbf{b} shows the \nicer\ 2–8 keV light curve for the same burst with the same time bin size. Panels \textbf{c} and \textbf{d} illustrate the light curves for burst 5. We present deadtime-corrected light curves from both mode 01 and mode 06 data. The broadband \nicer\ and \nustar\ spectra of burst 2 (light blue) and burst 5 (pink) are illustrated in panel \textbf{e}, while panels \textbf{f} and \textbf{g} display the corresponding residuals. The thick solid lines denote the best-fit BB+BB spectra, with the soft component represented by dotted lines and the hard component by dashed lines. }
    \label{fig:burst1904}
\end{figure*}

\subsection{Burst Spectroscopy} \label{sec:spectral_analysis}

\subsubsection{Broadband Spectra of Five Bursts}
To examine the spectral variability of bursts in detail, we initially extracted broadband spectra from bursts detected simultaneously by \nicer\ and \nustar.
Among all the bursts identified by \nustar, twenty are simultaneously observed with \nicer.
Five bright bursts, labeled as 1--5 in Figure \ref{fig:burst_evolution_all}a, are used for broadband spectral analysis.  
The most fluent burst (burst 5) that was simultaneously detected by both \nicer\ and \nustar\ occurred at $t=-1.83$ hours, which is close to the \nustar\ GTI boundary, with most photons captured in mode 06 data.
In addition, burst 5 is the only one that occurred after the intermediate flare, i.e., the sharp decrease of burst hardness ratio.
We plotted \nustar\ 3--79 keV and \nicer\ 2--8 keV light curves of the two brightest (Bursts 2 and 5) in Figure \ref{fig:burst1904} for reference. 
Burst 2 (Figures \ref{fig:burst1904}a and \ref{fig:burst1904}b) displays a major peak following the onset of the burst, and exhibits two small peaks on its tail.
In contrast, burst 5, characterized by its four peaks and a duration of approximately 2.5 s, is illustrated in Figures \ref{fig:burst1904}c and \ref{fig:burst1904}d. 
For burst 5, photons collected in mode 01 are not used for spectral analysis since most of the X-ray photons were captured in mode 06 observation.

We extracted X-ray spectra from all these five bright bursts detected by \nustar\ and \nicer\, and then jointly fitted the spectra for all five bursts.
To deal with interstellar absorption, we used the \texttt{tbabs} model with the corresponding interstellar medium abundance as described in \citet{WilmsAM2000} to determine the hydrogen column density ($N_{\textrm{H}}$). 
Initially, we allowed this parameter to vary freely and found that the $N_{\textrm{H}}$ values remained consistent across the five datasets, although the uncertainty of $N_{\rm{H}}$ can be very large for faint bursts. 
Therefore, we linked this parameter across all datasets for subsequent analyses.
Moreover, an additional constant component was introduced to account for cross-calibration between instruments and flux differences between \nicer\ and \nustar. 
Due to significant dead time issues with \nustar, we estimated the X-ray flux using \nicer\ data. 
We fit the data with two models: the first consists of two blackbodies (2BB, XSPEC model \texttt{bbodyrad}+\texttt{bbodyrad}) components, while the second is a cutoff power law (CPL, XSPEC model \texttt{cutoffpl}). 
The CPL model can be described as 
\[
A(E)=KE^{\Gamma}\exp\left(-E/E_f \right),
\]
where $\Gamma$ is the photon index, $K$ is the normalization constant, and $E_f$ is the e-folding energy of exponential rolloff. 
Both 2BB and CPL models adequately fit the spectra, where 2BB model performs slightly better than that of the CPL model regarding the goodness of fit. 
In the case of burst 1, we encountered a challenge with parameter insensitivity issues, making the parameters of the hard component difficult to constrain. 
Therefore, we determined the upper and lower limits of parameters using the steppar command.

The best-fit spectral parameters of these five bursts are shown in Table \ref{tab:bursrs_parameters}, where the spectra for Bursts 2 and 5 are shown in Figure \ref{fig:burst1904}e, with burst 2 peaking around 10 keV and burst 5 around 5 keV. 
The 2BB model result in an $N_{\textrm{H}}=(2.6\pm0.2)\times10^{22}$~cm$^{-2}$, consistent with previously determined values \citep[see, e.g.,][]{YounesKJ2017}, while the CPL model yielded a slightly larger $N_{\textrm{H}}=(3.6\pm0.1)\times10^{22}$~cm$^{-2}$.
Notably, the best-fit spectral parameters reveal that the temperature of both BB components in burst 5 is significantly lower than those in the other four bursts.
Regarding the CPL model, the burst 5 has $\Gamma=0.65\pm0.07$, which is significantly higher compared to the other bursts, and the e-folding energy $E_f$ is relatively lower. 
This aligns with the softening trend observed in Figure \ref{fig:burst_evolution_all}, indicating a spectral change occurs near the intermediate flare for both the soft and hard components of the spectra.

In this analysis, we also tried fitting the source and background simultaneously using the SCORPEON model and got consistent parameters. 
Additionally, fitting only the \nustar\ data produced spectral parameters that deviated from the broadband result. 
For instance, fitting burst 2 with a CPL model in \nustar\ data alone yielded $\Gamma=1.5\pm0.4$ and $E_f>15$ keV, which aligns more closely with the averaged burst emission in the pre-storm epoch (see Section \ref{sec:hid_evolution}). 
If we force $\Gamma$ at 1.5, the \nicer\ data shows significant residuals that cannot be eliminated by adjusting the SCORPEON parameters.
This discrepancy indicates that the true emission model is likely more complex than our current ones.

\begin{figure*}[t]
    \centering
    \includegraphics[width=0.99\textwidth]{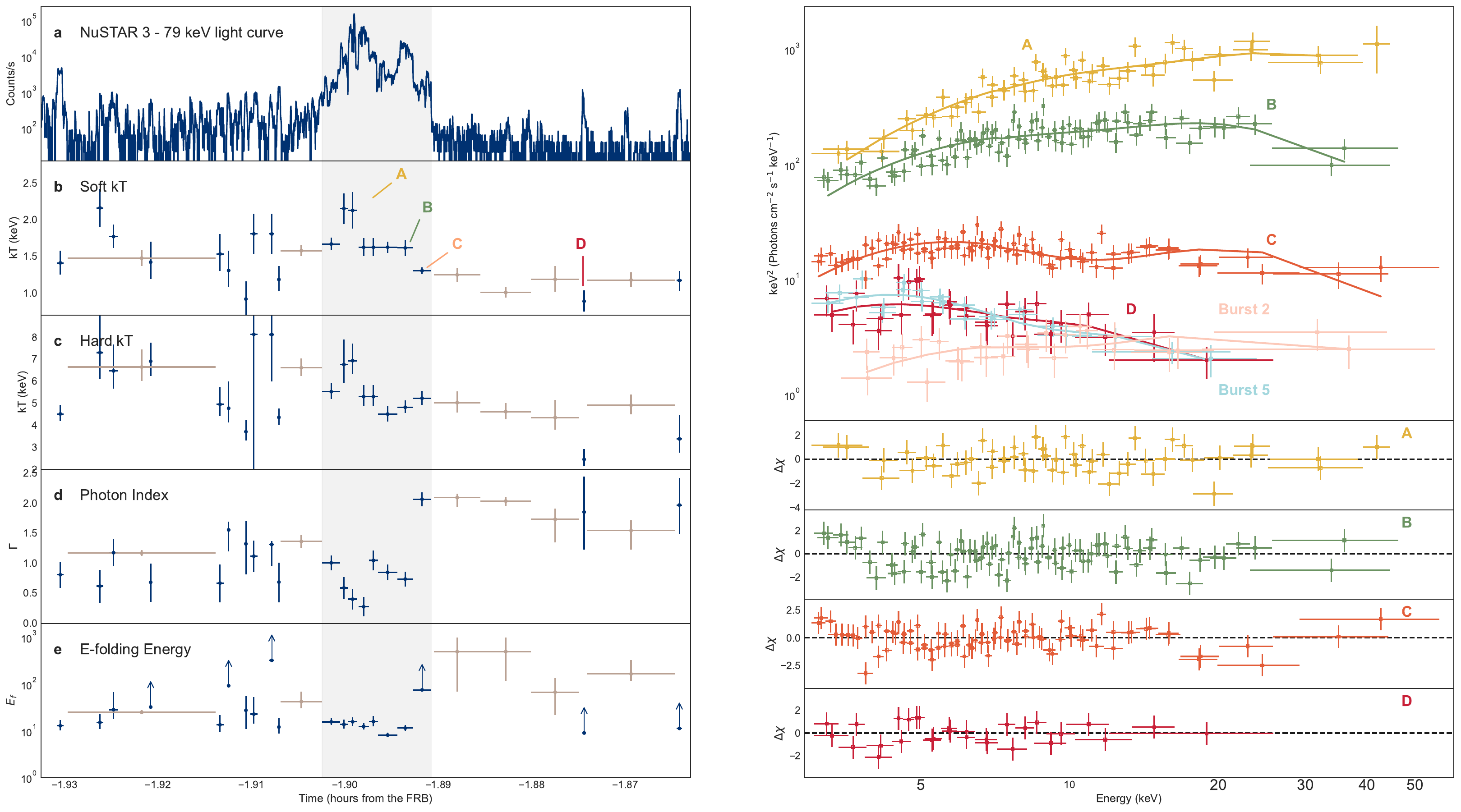}
    \caption{Spectral evolution close to the intermediate flare. Panel \textbf{a} shows the \nustar\ 3--79 keV light curve. Panels \textbf{b} to \textbf{e} show the spectral parameters, including the temperatures of the soft (panel \textbf{b}) and hard (panel \textbf{c}) BB components of the 2BB model, and the $\Gamma$ (panel \textbf{d}) and $E_f$ (panel \textbf{e}) of the CPL model. 
    The 40-s time interval of the intermediate flare, including the tail, is shaded in gray.
    The dark blue points represent the parameters of the bright short bursts and the parameters at different time intervals during the intermediate flare. 
    The light gray points represent the spectral parameters of the persistent emission, where the emission from peaks of faint bursts is removed.
    A, B, and C denote three segments of the intermediate flare, whereas D denotes a short burst soon after the intermediate flare.
    The \nustar\ spectra of A--D are shown in the right panel, where burst 2 and burst 5 are also plotted for comparison.}
    \label{fig:spectrum_huge_flare}
\end{figure*}

\subsubsection{\emph{NuSTAR} Spectra Close to the Intermediate Flare}\label{sec:intermediate_flare_softening}
We then carried out spectral analysis of the intermediate flare and bright X-ray bursts near the flare. 
They are different in time compared to the five bursts described in the previous section.
Due to these bursts falling outside the \nicer\ GTIs, we performed spectral analysis with \nustar\ data solely. 
Given \nustar's limitations in soft X-ray detection, we were unable to constrain the $N_{\rm{H}}$ value accurately.
Hence, we froze it at the best-fit values obtained from broadband spectroscopy as $2.6\times10^{22}$ cm$^{-2}$ for the 2BB model and $3.6\times10^{22}$ cm$^{-2}$ for the CPL model.

The spectral evolution near the intermediate flare is shown in Figure \ref{fig:spectrum_huge_flare}.
The \nustar\ 3--79 keV light curve, featuring this outburst peak and subsequent flux decline, is illustrated in Figure \ref{fig:spectrum_huge_flare}a. 
We note that the light curve reveals several nearly vertical luminosity changes. 
These discontinuities in luminosity occur at integer seconds of \nustar's mission elapsed time, possibly due to the effects of \nustar\ deadtime correction. 
Therefore, the flux and blackbody radius of the intermediate flare may not be reliable, and we did not perform any fitting on its tail to obtain possible flux-decreasing parameters.

The \nustar\ spectra of individual bursts and short-term persistent emission can be fit with both 2BB and CPL. 
Despite higher fluctuations in the burst spectra due to fewer collected photons, they exhibit similar behavior to the persistent emission. 
The best-fit temperature of the soft and hard BB components are shown in Figures \ref{fig:spectrum_huge_flare}b and \ref{fig:spectrum_huge_flare}c, whereas $\Gamma$ and $E_f$ of the CPL model are shown in \ref{fig:spectrum_huge_flare}d and \ref{fig:spectrum_huge_flare}e.
Before the intermediate flare, the persistent emission exhibited a soft component of $kT_{\rm{soft}}\approx1.5$~keV and a hard component of $kT_{\rm{hard}}\approx6.5$~keV. 
After the flare, these temperatures decreased to 1.1~keV and 4.5~keV, respectively. 
A rapid drop in temperature for both the soft and hard components can be clearly observed during the intermediate flare (Figure \ref{fig:spectrum_huge_flare}b and d). 
For the CPL model, $\Gamma$ was around 1 before the flare peak, jumping to $\Gamma\approx2$ after the peak.

We note that the \nustar\ data suffer from pile-up around the intermediate flare, as indicated by the GRADE and PRIOR parameter distributions in the event files \citep[see Appendix A in][for more details]{GrefenstetteGK2016}. 
Any interpretation that depends on precise spectral parameters or complex modeling may not be reliable. Nevertheless, the overall spectral evolutionary trends we observe remain robust despite these issues. 
\nicer\ did not capture the intermediate flare, and other fainter bursts are too weak to produce severe pile‐up.

To visualize the spectral evolution, we divided the intermediate flare into several segments, labeling three of them as A, B, and C. 
Additionally, we labeled a short X-ray burst occurring just after the flare as D. 
We plotted their X-ray spectra in the right panel of Figure \ref{fig:spectrum_huge_flare}.
The \nustar\ spectra of Bursts 2 and 5 are included for comparison. 
The spectral shape of segment B closely resembles that of burst 2, detected 15 hours before FRB 20221014A. 
Conversely, segment D, which is the first short burst after the intermediate flare, is significantly softer compared to bursts before the intermediate flare and shares a similar spectral profile with burst 5 in Figure \ref{fig:burst1904}. 
These observations indicate a rapid spectral evolution within a timescale of 80~s. 

We also performed a spectral analysis of the entire intermediate flare. During the entire exposure, our data could not be adequately described using the 2BB model due to rapid spectral variations. 
Interestingly, within the \nustar\ band, the overall spectrum can be described with a CPL. 
The best-fit parameters are $\Gamma=1.28\pm0.04$ and $E_f=22\pm2$ keV.  
The time-integrated spectral flux yielded a total energy radiated by this intermediate flare as $(6.3\pm0.2)\times10^{40}$~erg using the best-fit model.
This energy output is much lower than that of magnetars' known giant flares, although it might be underestimated due to deadtime issues.

\begin{figure*}[t]
    \centering
    \includegraphics[width=0.99\textwidth]{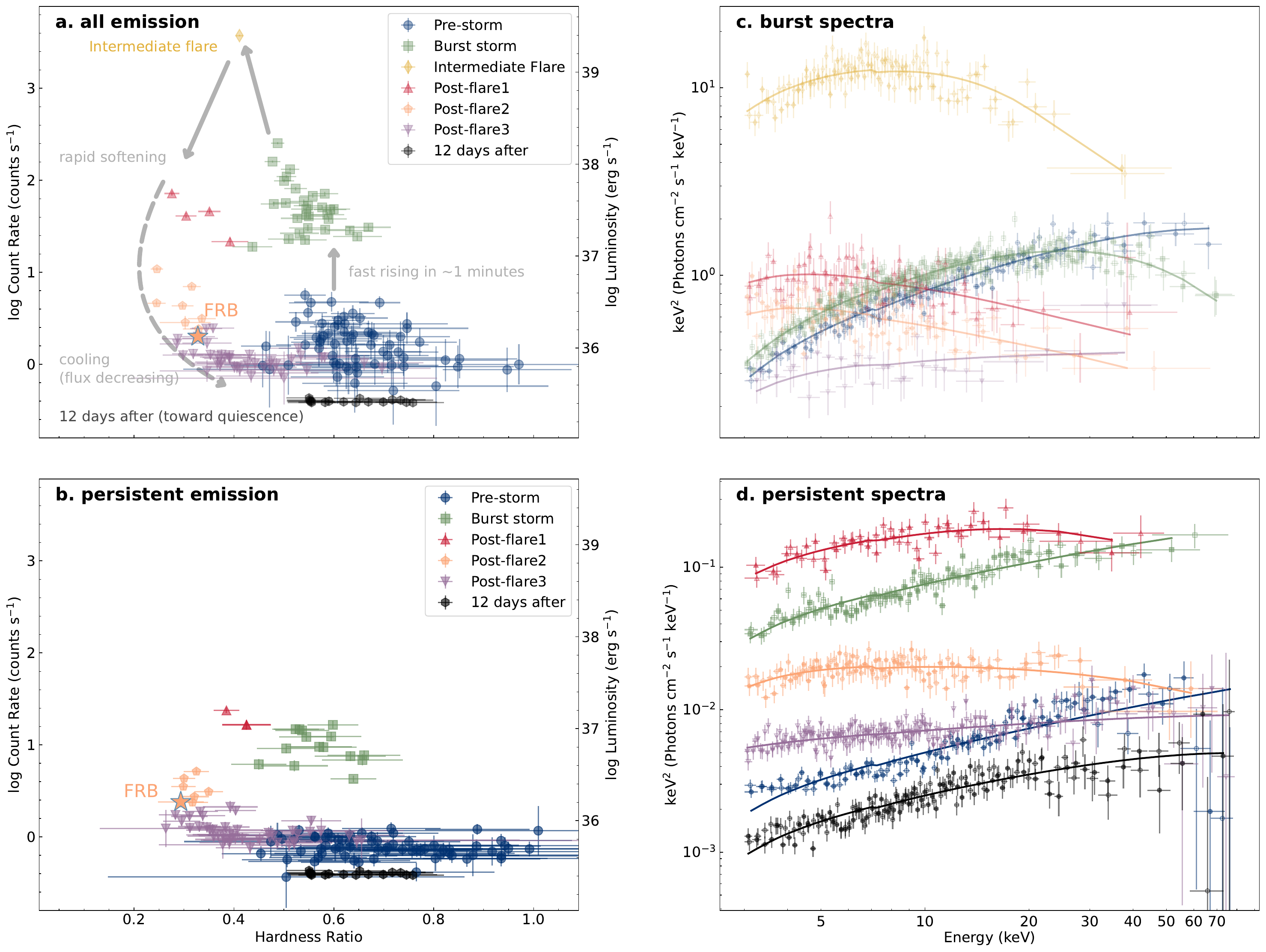}
    \caption{\textbf{a.} Hardness ratio versus count rate for all (persistent and burst) emission observed with \nustar. Blue circles and green squares represent data from the pre-storm and burst-storm epochs, respectively. The post-flare epoch is divided into three segments: post-flare 1 (red triangles), post-flare 2 (orange pentagons), and post-flare 3 (purple downward triangles). The time intervals for these epochs are defined in Figure \ref{fig:burst_evolution_all}. Data taken from \nustar\ ObsID 80802317004 are shown as gray hexagons. The orange star marks the time bin 10 minutes before FRB 20221014A. Gray arrows indicate the evolutionary track of \src\ during the \nustar\ observation. Pre-storm and post-flare data points were derived from 512-s binned light curves, while burst-storm data points were derived from 64-s binning. Data taken 12 days after FRB 20221014A were binned according to the \nustar\ orbit (96.8 minutes). The luminosity is estimated using the same assumptions applied when computing the fluence in Figure \ref{fig:burst_evolution_all} except for the intermediate flare as we have a spectral fit. \textbf{b.} Same as panel \textbf{a} but only for persistent emission. The number of data points in the burst-storm and post-flare 1 epochs is fewer than in panel \textbf{a} because we use a bin size of 128 s.  \textbf{c.} Accumulated \nustar\ 3--79 keV spectra of bursts detected in each epoch defined in panel \textbf{a}. \textbf{d.} Accumulated \nustar\ 3--79 keV spectra of persistent emission for each epoch defined in panel  \textbf{a}. The colors and symbols of the spectra are consistent with those defined in panel \textbf{a}, with burst spectra represented in more transparent colors than those of the persistent spectra. Filled and open circles represent data collected by FPMA and FPMB, respectively.}
    \label{fig:persistent_hid}
\end{figure*}

\subsection{Evolutionary Pattern on the Hardness-Intensity Diagram}\label{sec:hid_evolution}
In the spectral analysis, we found that spectral behavior and evolutionary patterns of the persistent and burst emission may share the same properties.
Based on this, we further examined the hardness-intensity relationship of SGR 1935+2154 using all available X-ray photons in the 3--79 keV range obtained with \nustar.
We divided the entire \nustar\ evolution into five epochs:
\begin{itemize}
    \item pre-storm: from the beginning of the \nustar\ observation to the onset of burst storm (2.5 hr before FRB 20221014A).
    \item burst storm: from the onset of the burst storm to one minute before the intermediate flare (1.9 hr before FRB 20221014A).
    \item post flare 1: right after the intermediate flare to the end of this GTI.
    \item post flare 2: the next two GTIs after the intermediate flare. FRB 20221014A occurred 10 minutes after this segment. 
    \item post flare 3: after FRB 20221014A. 
\end{itemize}
The time intervals of these epochs are shown in Figure \ref{fig:burst_evolution_all}. 
For pre-storm, post-flare 2, and post-flare 3 epochs, we calculated the hardness ratio using the 512-s binned light curve and plotted the hardness-intensity diagram.
The hardness ratio is obtained from 64-s binned light curves for burst-storm and post-flare 1 epochs owing to their high count rate and short duration. 
Finally, the data taken 12 days after FRB 20221014A (\nustar\ obsid 80802317004) is binned according to the \nustar\ orbital period. 

The hardness-intensity diagram (Figure \ref{fig:persistent_hid}a) exhibits an intriguing evolutionary pattern, suggesting an association with magnetospheric state and FRB generation.
During the pre-storm epoch, the hardness ratio ranged mostly between 0.5 and 1.0, with an average value of 0.7, and the average count rate remained low at around less than 10 count s$^{-1}$. 
At $t=-2.5$ hours, a short-term outburst occurred, which caused the persistent count rate to increase rapidly within a timescale of 1 minute.
A burst storm occurred simultaneously, and the burst occurrence rate increased by an order of magnitude. 
Despite these energetic events, the hardness ratio remained within the range of 0.5 to 0.7, slightly lower than the pre-storm epochs. 
A possible negative correlation between the count rate and the hardness ratio was also observed with a linear correlation coefficient of $-0.37$.
This correlation is only marginally detected, as a bootstrap simulation revealed that the standard deviation of the linear correlation coefficient in the re-sampled data sets is 0.2.
Later, at $t=-1.9$ hours, an intermediate flare took place, accompanied by rapid spectral softening with a timescale of about 40 s. 
This intermediate flare and a few bright bursts during the burst storm exceed the Eddington luminosity. 
Soon after this intermediate flare, the hardness ratio significantly dropped to $\lesssim 0.4$. 
Following this, the flux gradually decreased, and the hardness ratio slowly recovered to around 0.6.
Although the X-ray burst associated with FRB 20221014A was not captured during this observation, the closest time bin of the light curve is marked with an orange star in Figure~\ref{fig:persistent_hid}. 
The evolution of the entire process traces a counterclockwise pattern in the hardness-intensity diagram.
We performed the same analysis on the persistent emission alone and observed a consistent pattern, though at a lower count rate.

\begin{table*}
 \caption{Best-fit parameters of X-ray spectra in Figure \ref{fig:persistent_hid}. }
 \label{tab:spectral_parameters_epochs}
  \begin{tabular}{c|cccc|cccc}
 \hline
 & \multicolumn{4}{c}{Persistent Emission} & \multicolumn{4}{c}{Burst Emission} \\
 \hline
 & $\Gamma$& $E_f$ (keV)& stat/dof& Prob. & $\Gamma$& $E_f$ (keV)& stat/dof & Prob.\\
 \hline
 pre-storm& $1.44\pm0.02$& $>430$& 1069.9/1000& 0.04 & $1.29\pm0.04$& $90_{-20}^{+30}$& 993.8/1002 & 0.08\\
 burst storm& $1.49_{-0.04}^{+0.01}$& $>205$& 684.2/682& 0.19 & $1.17\pm0.02$& $30\pm2$& 1473.6/1424& 0.008\\
 intermediate flare&  ...&  ...&  ...& ... & $1.28\pm0.04$& $22\pm2$& 865.5/861 & 0.07\\
 post-flare 1& $1.5\pm0.1$& $31_{-8}^{+15}$& 351.5/369 & 0.47 & $2.26_{-0.14}^{+0.10}$& $>38$& 332.5/352& 0.7\\
 post-flare 2& $1.92\pm0.08$& $80_{-30}^{+100}$& 465.5/530& 0.92 & $2.37_{-0.05}^{+0.01}$& $>170$& 362.1/364& 0.39\\
 post-flare 3& $1.90\pm0.02$& $>370$& 755.2/822& 0.83 & $1.90\pm0.08$& $>210$& 100.6/105& 0.60\\
 12-days after& $1.42\pm0.05$& $120_{-50}^{+160}$& 949.6/867& 0.002& ...& ...& ...&...\\
\hline
\end{tabular}
\end{table*}

The averaged X-ray spectra of the burst and persistent emission during these epochs are shown in Figures \ref{fig:persistent_hid}c and \ref{fig:persistent_hid}d, respectively. 
In our burst spectral analysis, we used the corresponding persistent spectrum as the background except for the intermediate flare. 
We also fitted the combined burst and persistent spectra directly and obtained fully consistent results.
Although the broadband spectra of individual bursts can be modeled slightly better using 2BB compared to CPL, as described in previous sections, the accumulated burst spectra are much better described with a single CPL, except for the pre-storm persistent emission. 
Therefore, we use this model to fit the data and trace the evolution of the spectral parameters (see Table \ref{tab:spectral_parameters_epochs} for best-fit spectral parameters).

Before the burst storm, the averaged persistent emission is best fitted by a soft blackbody plus a hard power law with a possible high-energy cutoff, as shown in \citet{HuNE2024}. 
If we use the CPL to describe the spectrum, the photon index is $\Gamma = 1.44 \pm 0.02$, and the e-folding energy cannot be well constrained (blue spectrum in Figure \ref{fig:persistent_hid}d). 
During the same epoch, the accumulated burst emission can be modeled with a CPL with $\Gamma = 1.29 \pm 0.04$ and $E_f=90^{+30}_{-20}$ keV. 
The spectral shapes of the persistent and burst emission are similar, although the burst emission is slightly more curved than the persistent spectrum and no signature of the soft blackbody tail can be seen (transparent blue spectrum in Figure \ref{fig:persistent_hid}c).

During the burst-storm epoch, the averaged persistent emission can be described with a CPL with $\Gamma = 1.49^{+0.01}_{-0.04}$, and the $E_f$ still cannot be well constrained. 
This suggests that the persistent emission is slightly softened compared to the pre-storm persistent emission (green spectrum in Figure \ref{fig:persistent_hid}d). 
The burst emission is more significantly curved, with $\Gamma = 1.17 \pm 0.02$ and $E{\rm{cut}} = 30 \pm 2$ keV. 
The burst spectrum shows significant curvature, peaking at around 30 keV (transparent green spectrum in Figure \ref{fig:persistent_hid}c).

The intermediate flare showed a significant softening, as described in Section \ref{sec:intermediate_flare_softening}. 
Interestingly, the accumulated spectrum can still be well described with a CPL with $\Gamma = 1.28 \pm 0.04$ and $E_f = 22 \pm 2$ keV. 
The spectral curvature is similar to that of the burst emission during the burst-storm epoch, but the peak migrates from approximately 30 keV to below 20 keV.

Soon after the flare (post-flare 1), the curvature of the persistent spectrum is clearly seen, and the best-fit spectral parameters are $\Gamma = 1.5 \pm 0.1$ and $E_f = 31^{+15}_{-8}$ keV (red spectrum in Figure \ref{fig:persistent_hid}d). 
At the same time, the burst spectrum also shows significant curvature, but the peak has shifted to around 5 keV, and the power-law tail is much softer than in any epochs before the intermediate flare. 
Therefore, the best-fit CPL model yields a large $\Gamma = 2.26_{-0.14}^{+0.10}$ and an ill-constrained $E_f > 38$ keV (transparent red spectrum in Figure \ref{fig:persistent_hid}c).

In the second GTI after the intermediate flare (post-flare 2, just  before FRB 20221014A), the persistent emission still shows significant curvature, and the best-fit spectral parameters are $\Gamma = 1.92 \pm 0.08$ and $E_f = 80_{-30}^{+100}$ keV (orange spectrum in Figure \ref{fig:persistent_hid}d). 
Interestingly, the emission above 40 keV has decreased to the pre-storm level; however, the soft ($<40$ keV) part remains several times higher than that of the pre-storm level. 
At the same time, the burst emission becomes even softer than that of the post-flare 1 epoch, with $\Gamma = 2.37_{-0.05}^{+0.01}$ and $E_f > 170$ keV (transparent orange spectrum in Figure \ref{fig:persistent_hid}c).

After FRB 20221014A (post-flare 3), both the persistent and burst emission become harder again, and their spectral parameters are consistent with each other. The power-law tail of the persistent emission above 20 keV has become fainter than that of the pre-storm emission, although the soft X-ray emission remains higher.

The \nustar\ data collected 12 days after this event show no burst emission, and the spectrum of the persistent emission has evolved back to the same shape as the pre-storm spectrum, but the flux has dropped by a factor of two, as reported in \citet{HuNE2024}.

\begin{figure*}[t]
    \centering
    \includegraphics[width=0.85\textwidth]{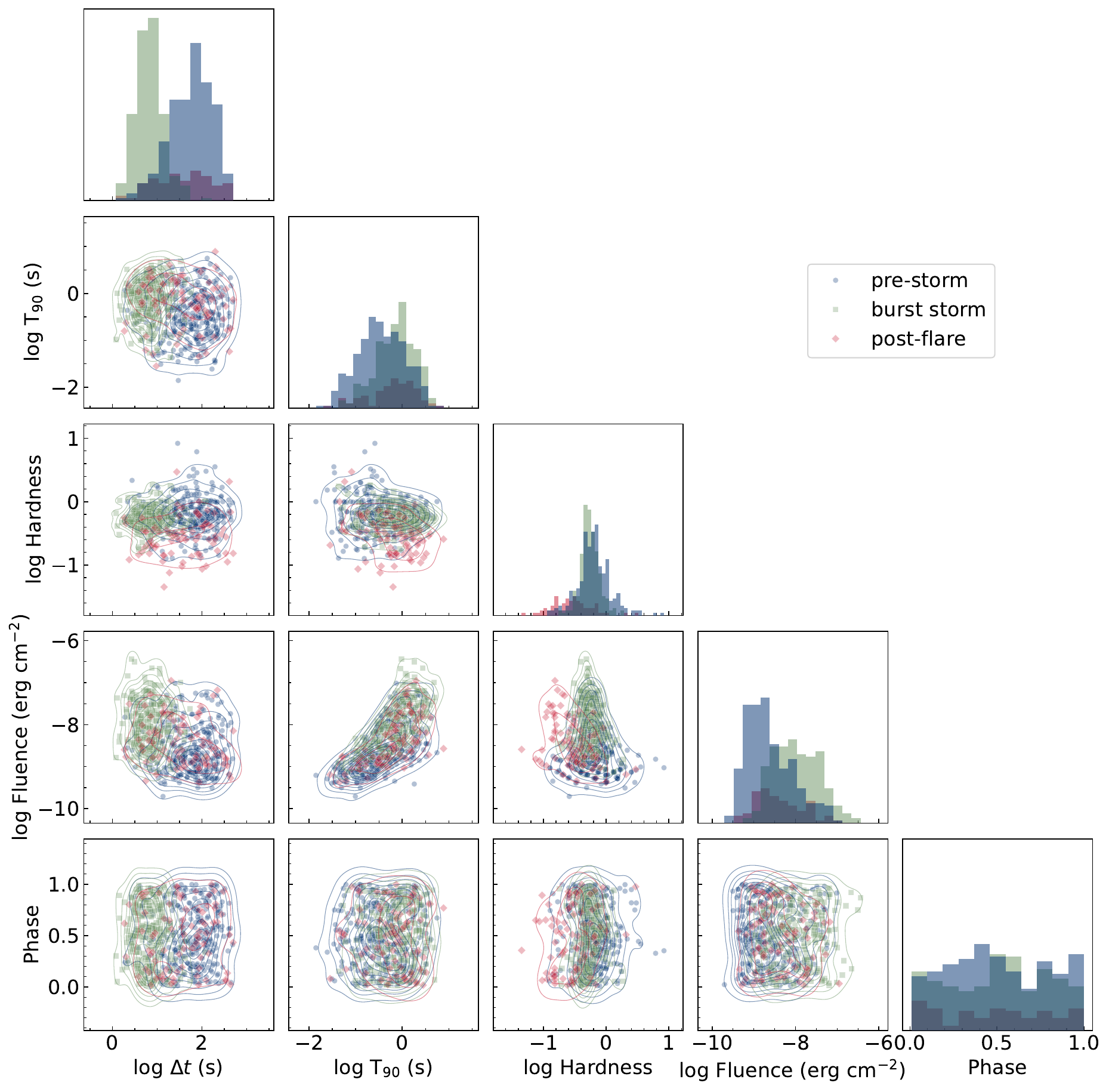}
    \caption{Relationship between five parameters of \nustar\ bursts: waiting time ($\Delta t$), T90, burst fluence, hardness ratio, and pulse phase. Blue circles, green squares, and red diamonds represent bursts detected in pre-storm, burst-storm, and post-flare epochs, respectively.  The histogram of individual parameters is plotted on the top panels of each column. The two-dimensional KDE overlays the data points on each panel.}
    \label{fig:burst_correlation}
\end{figure*}

\subsection{\nustar\ Burst Parameter Statistics}
We further explore the statistics and the relationship between five parameters of bursts: the $T_{90}$, waiting time ($\Delta t$), burst fluence, hardness ratio, and pulse phase. 
Given that individual bursts can span several rotational cycles of the magnetar, we determined peak times by binning the light curve into 0.01-s intervals and selecting the highest count rate within T90 as the peak. 
The pulse phase for each burst is determined based on the burst's peak time, utilizing the double-glitch ephemeris derived in \citet{HuNE2024}.
We calculated the $\Delta t$ for each burst, defined as the interval between a burst and the one preceding it. 
The first burst of each \nustar\ GTI was excluded from the analysis, as it is uncertain whether any bursts occurred during the data gap.
Following the time epochs defined in Section~\ref{sec:hid_evolution}, we classified \nustar\ bursts into three groups: pre-storm, burst-storm, and post-flare groups. 
The relationships between these parameters, along with two-dimensional kernel density estimations (KDE) contours, are shown in Figure \ref{fig:burst_correlation}. 
Histograms for each parameter are displayed at the top of each column, with bursts from the pre-storm, burst-storm, and post-flare epochs individually color-coded.

This plot reveals several intriguing features, especially regarding the distribution and correlation related to $\Delta t$ and hardness ratio. 
First, the $\Delta t$ distributions of pre-storm and burst-storm bursts differ significantly. 
Second, while the phase distribution of bursts shows fluctuations, it generally remains consistent with a uniform distribution. 
However, the pulse phase distribution of the burst hardness ratio exhibits a curved pattern, where the hardness ratio reaches maximum at phase 0.5. 
This may provide hints about the connection between burst emission and persistent emission. 
Finally, the hardness ratios of bursts detected during the pre-storm and burst-storm epochs are similar; however, the post-flare bursts are systematically softer than the other two groups. 
This can be seen in both the histogram of hardness ratio and the fluence-hardness panel of Figure \ref{fig:burst_correlation} as well as in Figure~\ref{fig:persistent_hid}.
We further investigate and discuss these three points in the following sections.

\subsubsection{Distribution of $T_{90}$ and fluence}

The overall distribution of $T_{90}$ has a mean value of 0.87 s. Bursts detected during the burst storm have a higher average $T_{90}$ of 1.07 s compared to pre-storm bursts, which have an average $T_{90}$ of 0.66 s. 
The difference between these two groups is significant, with a p-value of $4.2\times10^{-11}$ (approximately 6.6$\sigma$) as determined by a Kolmogorov–Smirnov (KS) test.
Post-flare bursts exhibit a similar distribution to burst-storm bursts, also averaging $T_{90} = 1.07$ s.

The burst fluence across these three groups also shows distinct distributions. 
The bursts detected in pre-storm and burst-storm epochs have mean fluences of $6.1 \times 10^{-9}$ erg cm$^{-2}$ and $2.4 \times 10^{-8}$ erg cm$^{-2}$, respectively. 
A KS test indicates these two sets are drawn from different distributions at 11$\sigma$ level. 
The mean fluence of post-flare bursts is $9.5 \times 10^{-9}$ erg cm$^{-2}$.

The differences in fluence and $T_{90}$ between the pre-storm and burst-storm epochs likely result from faint, short-duration bursts being buried by the enhanced persistent emission during the burst storm, making them undetectable. 
This is also reflected in the burst fluence distribution, where low-fluence bursts are absent during the burst-storm epoch. 
Post-flare bursts may represent a cooling phase, exhibiting properties that fall between the pre-storm and burst-storm epochs.

The observed $T_{90}$ values are close to the average value of 0.84 s found during the 2020 event \citep{Younes2020}, which is among the longest within the magnetar burst family, but shorter than those observed with \emph{Fermi} GBM during with the same 2022 outburst \citep{RehanI2025}. 
\citet{Younes2020} suggested that the relatively long durations are due to \nicer's low background compared to Fermi GBM or RXTE. 
This effect is even more pronounced in our study, as \nustar\ offers better spatial resolution than \nicer.

This difference in $T_{90}$ estimates between instruments may also be driven by the spectral behavior of individual bursts and the energy band of individual instruments. 
Additionally, it can be challenging to determine whether a series of consecutive peaks represents multiple distinct bursts or a single, long burst with multiple peaks. 
Recognizing such events as a single burst with several peaks may contribute to longer average $T_{90}$ estimates.

\begin{figure*}[t]
\includegraphics[width=0.99\textwidth]{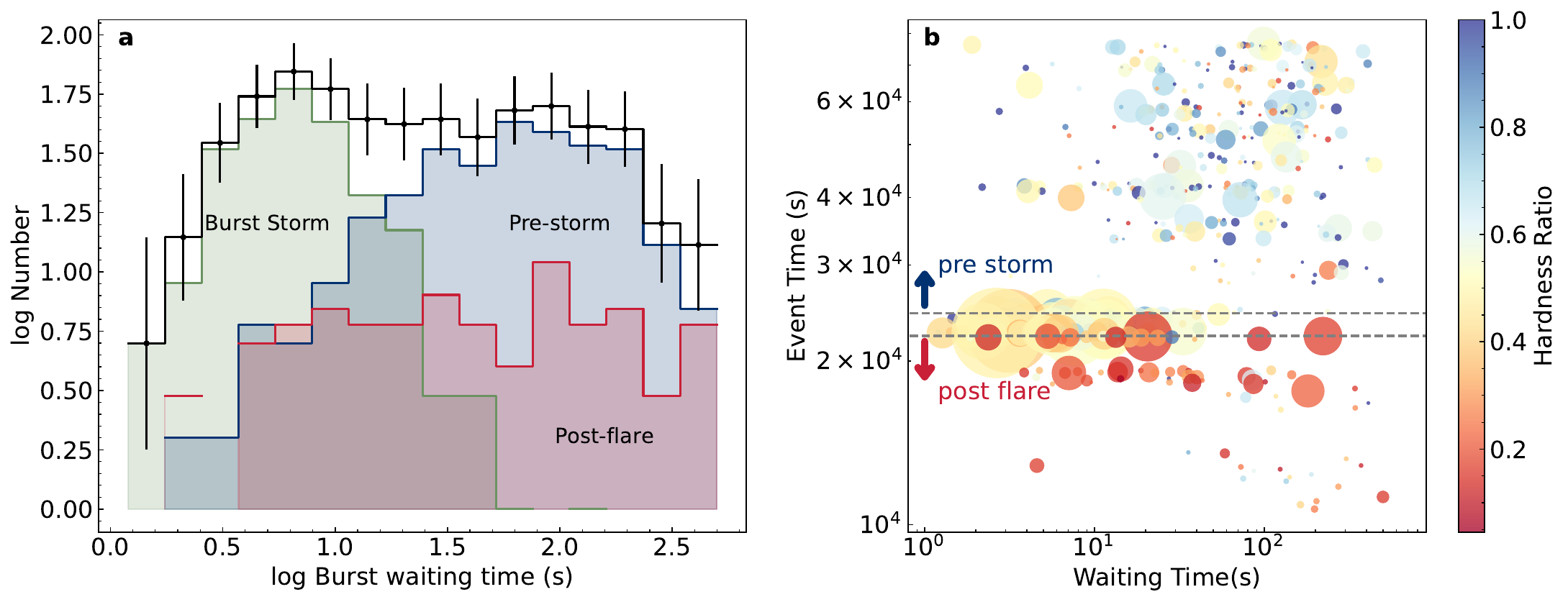}
\caption{\textbf{a} The black histogram shows the $\Delta t$ distribution of all bursts detected with \nustar. Bursts detected during the pre-storm, burst-storm, and post-flare epochs are filled with blue, green, and red, respectively. \textbf{b} Burst $\Delta t$ versus event time. Time zero is defined as the last burst detected before the second glitch, and the event time is counted backward from this epoch. The hardness ratio of each burst is color-coded, where the bluer color represents a harder spectrum and the redder color denotes a softer spectrum. The size of the bubble denotes the burst fluence, whereas larger bubbles represent bursts with higher fluences.  \label{fig:burst_waiting_time}}
\end{figure*}

\subsubsection{Burst Waiting Time}

The $\Delta t$ distributions during the pre-storm, burst-storm, and post-flare epochs are notably different from one another (see Figure \ref{fig:burst_correlation}), providing key insights into the magnetar's activity. 
In previous studies, the $\Delta t$ of short X-ray bursts from magnetars has been found to follow a lognormal distribution (e.g., SGR 1900+14, \citealt{GogusWK1999}; SGR1806$-$20, \citealt{GogusWK2000}; 1E 2259+586, \citealt{Gavriil2004}; and the 2020 burst from SGR 1935+2154, \citealt{Younes2020}). 
However, the overall $\Delta t$ distribution for this 2022 outburst is double-peaked, which is unlike previous observations. 
This can be attributed to the significantly different burst occurrence rates between the pre-storm epoch (which remained burst-active) and the burst-storm epoch.

During the pre-storm epoch, the burst occurrence rate was relatively steady at 0.6 bursts per minute. 
The bursts in this phase follow a lognormal $\Delta t$ distribution with a mean value of 93 s, suggesting a self-organized process (see Figure \ref{fig:burst_waiting_time}). 
In contrast, during the burst-storm epoch, the burst occurrence rate increased quickly to 7.2 bursts per minute, much higher than the pre-storm value. 
Interestingly, the $\Delta t$ distribution remains lognormal-like but shifts to a much shorter regime with a mean value of 8.8 s. 
This value is significantly shorter than that of 1E 2259+586 (46.7 s, \citealt{Gavriil2004}), SGR 1900+14 (49 s, \citealt{GogusWK1999}), and SGR 1806$-$20 (103 s, \citealt{GogusWK2000}), but comparable to the 2020 burst storm (2.1 s, \citealt{Younes2020}). 
This rapid change in occurrence rate leads to the double-peaked structure in the $\Delta t$ histogram. 
After the intermediate flare, the burst occurrence rate dropped significantly, but unlike in previous phases, the post-flare bursts do not follow a lognormal distribution. 
Instead, they appear uniformly distributed in the logarithm of $\Delta t$, with a mean of 87 s, indicating a power-law decay in burst occurrence. This is similar log-uniform behavior to that seen in 1E 2259+586 \citep{Gavriil2004} and some episodes of FRB 121102 \citep{Wadiasingh2019}.

To further explore the decay of the burst occurrence, we plotted $\Delta t$ against burst peak time (Figure \ref{fig:burst_waiting_time}b), with color coding representing hardness ratio: bluer points correspond to harder bursts, while redder points represent softer bursts. 
Fluence is indicated by the size of the circles, with larger circles representing higher-fluence bursts. 
Unlike previous studies  \citep[e.g.,][]{Gavriil2004}, we set the time zero epoch at a burst occurring before the second glitch, calculating each burst's occurrence time backward from this point.

Interestingly, before the intermediate flare, the burst occurrence rate did not exhibit a significant decline, and no power-law decay was observed. 
After the intermediate flare, the relationship between $\Delta t$ and burst peak time ($t_p$) could not be fit with a simple power law, indicating a different decay behavior compared to bursts from other sources. 
Moreover, the overall hardness ratio shows a clear decreasing trend, consistent with the spectral softening observed in both the bursts and persistent emission, as discussed in previous sections.

\subsubsection{Hardness-Fluence Relationship}
The nature of magnetar burst emission can be inferred from the relationship between the spectral hardness and the fluence \citep[see, e.g.,][]{GogusKW2001, Gavriil2004, IsraelRM2008, YounesKv2014}. 
In the 2022 outburst of \src, we find that the burst hardness–fluence diagram exhibits a similar evolutionary pattern to the hardness–intensity diagram of the combined burst and persistent emission.
Although the fluences of bursts detected during the burst storm are slightly higher than those detected in the pre-storm epoch, no significant gap is observed between these two groups. 
Moreover, the post-flare bursts are significantly softer than bursts detected in the other two epochs, forming a distinct group in the hardness–fluence diagram. 
This behavior has also been observed in the burst parameter evolution (Figure \ref{fig:burst_evolution_all}) and the averaged burst spectroscopy across different epochs.

These results may indicate that apparent positive or negative correlations observed in previous studies could represent behaviors at different evolutionary stages when the time coverage is insufficient. 
Of course, not all magnetar outburst events exhibit such a short-term evolution that demonstrates the sequence of burst activity: burst active, burst storm, intermediate (or giant) flare, and post-flare cooling epochs. 
Therefore, high-cadence broadband monitoring of future magnetar outbursts would be needed to reveal unbiased correlations between these two parameters.

\begin{figure}
    \centering
    \includegraphics[width=0.49\textwidth]{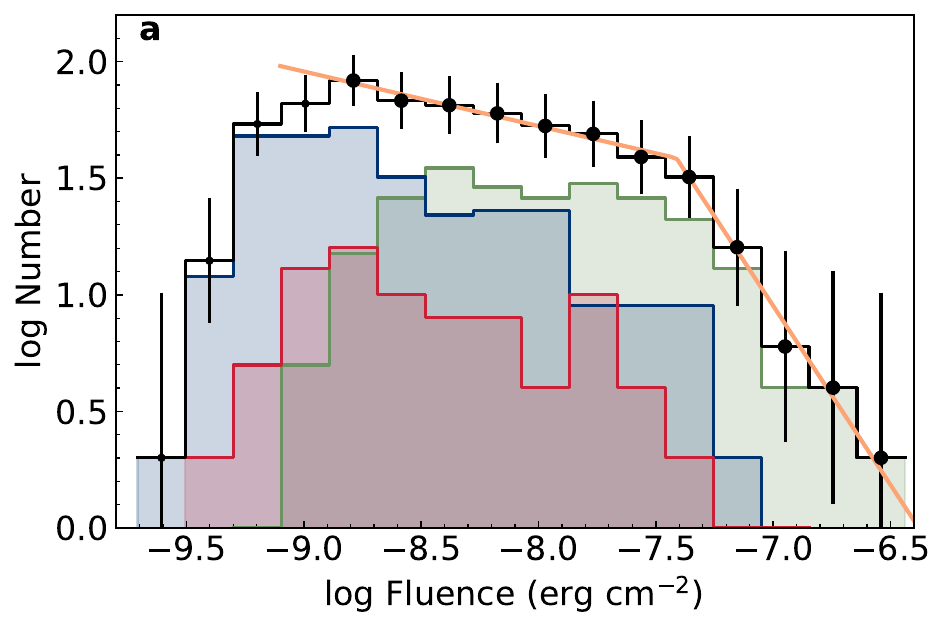}
    \includegraphics[width=0.49\textwidth]{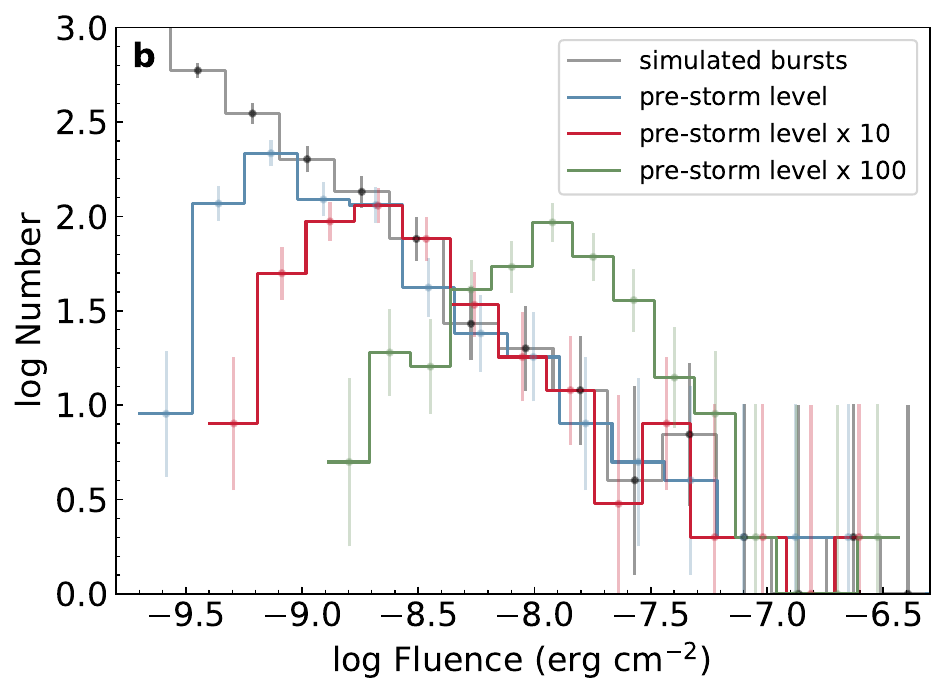}
    \caption{\textbf{a}. Fluence distribution of bursts detected with \nustar\ from \src. The filled dots represent bursts detected with full sensitivity and are used for fitting. The number of bursts with fluence lower than $10^{-9}$ erg cm$^{-2}$ could be underestimated due to sensitivity drop-off. The orange solid line represents a broken power-law fit. The definitions of the color-filled areas are the same as those in Figure \ref{fig:burst_waiting_time}a. \textbf{b}. Simulated burst samples and observed ones at various persistent emission levels. While extremely high and variable persistent emission can distort the overall distribution, the slope of the high-fluence tail remains largely unaffected. }
    \label{fig:burst_logN_logS}
\end{figure}

\subsubsection{Fluence Distribution of Bursts}
The fluence distribution of all short bursts, binned on a logarithmic scale, is shown in Figure \ref{fig:burst_logN_logS}a, along with histograms for the pre-storm, burst-storm, and post-flare epochs. The turnover at fluences lower than approximately $10^{-9}$ erg cm$^{-2}$ is due to the inability to recover bursts with lower fluences. Unlike other magnetars and the 2020 burst storm, the distribution above this turnover cannot be fitted with a simple power law. Instead, a broken power law describes the distribution very well. The turning point is located at a fluence of $10^{-7.4}$ erg cm$^{-2}$ ($4 \times 10^{-8}$ erg cm$^{-2}$). Below this turning point, the power-law index is $-0.26 \pm 0.07$, while above it, the index is $-1.5 \pm 0.3$. If we instead fit the data points with a single power law, the slope would be $-0.68 \pm 0.08$.

The pre-storm and post-flare bursts show a power-law-like distribution for fluences higher than $1.5 \times 10^{-9}$ erg cm$^{-2}$. The best-fit power-law indices are $-0.8 \pm 0.2$ for pre-storm bursts and $-0.6 \pm 0.2$ for post-flare bursts. These values are much flatter than the index detected during the burst storm but are consistent with the 2020 burst storm and several other magnetars \citep{Gavriil2004, Younes2020, YounesHB2022}. On the other hand, the broken power-law shape of the distribution could be caused by the difficulty in detecting bursts fainter than $4 \times 10^{-8}$ erg cm$^{-2}$ due to the burst activity and high persistent emission. If we fit the distribution of the burst storm above this fluence, the resulting power-law index is $-1.2 \pm 0.3$.  

To test the bias of enhanced persistent emission on burst detection, we created a set of 2000 simulated bursts and placed them into different levels of simulated unpulsed persistent emission (for the pulsed case, see Section \ref{sec:burstphases}). The gray histogram in Figure \ref{fig:burst_logN_logS}b shows the injected burst samples with a power-law fluence distribution with an index of $-1.0$ and a lognormal T90 distribution. We tested three persistent emission levels: (1) the same as the pre-storm level, (2) ten times the pre-storm level, and (3) one hundred times the pre-storm level (roughly matching the flare peak). We found that in cases (1) and (2), the high-fluence power-law tail remains unchanged, and contamination leads to a noticeable low-fluence cutoff. On the other hand, for case (3), the burst fluence can be overestimated due to the enhanced persistent emission, although the slope of the high-fluence tail remains largely the same. This simulation suggests that the observed flat top in the burst-storm fluence distribution, as well as the overall flat power-law index below $4 \times 10^{-8}$ erg cm$^{-2}$, could be caused by the strong and highly-variable persistent emission of the mini-outburst.

\begin{figure}
    \centering
    \includegraphics[width=0.49\textwidth]{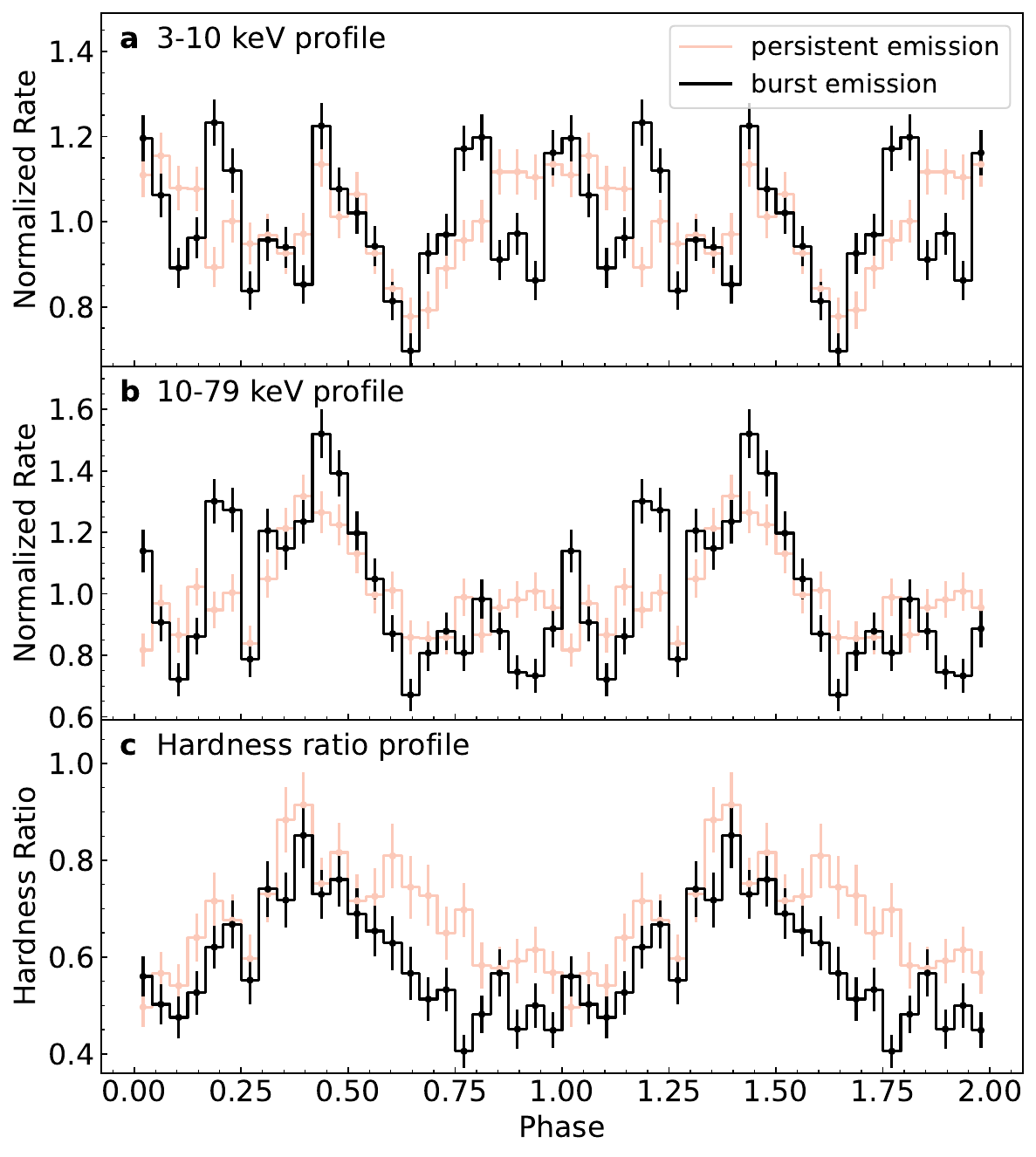}
    \includegraphics[width=0.495\textwidth]{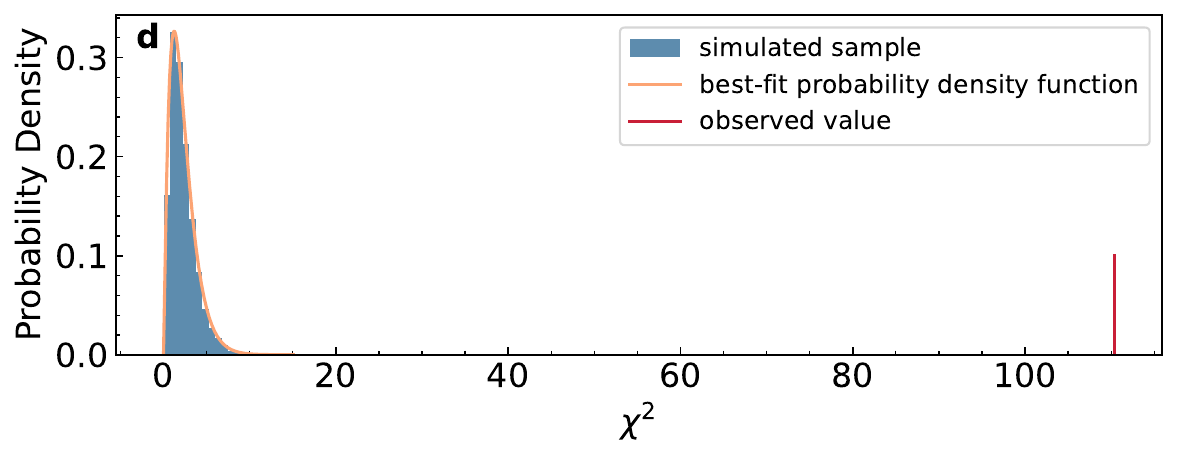}
    \caption{Folded light curves and hardness ratios from the pre-storm epoch. Pink profiles represent persistent emission, while black profiles show burst emission in the same energy range. Panels \textbf{a} and \textbf{b} represent the distributions of persistent and burst photons in the 3 -- 10 keV and 10 -- 79 keV bands, respectively. Panel \textbf{c} shows the corresponding hardness ratios. Panel \textbf{d} shows the resulting $\chi^2$ distribution from simulated samples and the observed $\chi^2$ value, suggesting that the observed phase dependence of the burst hardness ratio is unlikely caused by a bias from contamination from the persistent emission.}
    \label{fig:burst_phase_distribution}
\end{figure}

\subsubsection{Spin-Phase Dependence of Burst Photons}
\label{sec:burstphases}

An intriguing finding from the parameter correlations is the curved pattern observed in the hardness-phase relationship, where the hardness ratio of bursts appears to peak at phase 0.5.
To investigate the dependence on the spin phase, we collected all X-ray photons detected with \nustar\ during the bursts and calculated their folded light curve and folded hardness ratio.
The post-flare bursts are excluded in this analysis because their hardness ratio significantly changed and the number of bursts is too few to have significant statistics. 
The burst storm epoch is also excluded from this analysis since the enhanced persistent emission could heavily contaminate the burst detection. 
We plotted the folded light curve of persistent emission in the 3--79 keV ranges, along with the folded light curves of burst photons collected during the pre-storm epoch (Figure \ref{fig:burst_phase_distribution}a and \ref{fig:burst_phase_distribution}b). 
The \nustar\ 3--79 keV light curve reveals a double-peaked profile, with the primary soft X-ray peak at the ephemeris's fiducial point (phase 0) and a secondary hard peak around phase 0.5 (Figure \ref{fig:burst_phase_distribution}a), as outlined in \citet{HuNE2024}.

Alternatively, the phase distribution of burst photons does not align with the folded light curve of the persistent emission. 
Compared to the folded light curve, the folded hardness ratio demonstrates a clearer alignment between burst emission and persistent emission. 
The hardness ratio for persistent emission shows a broad, sinusoidal-like profile. 
Interestingly, the hardness ratio profile of burst emission shows a similar pattern, as illustrated in Figure \ref{fig:burst_phase_distribution}b, aligning with but a bit softer than the hardness ratio profile of the persistent emission. 

To check whether this phase dependence is real or simply due to the bias posed by the pulsations of persistent emission, we performed a simulation using the same number of bursts and the same fluence and T90 distributions as in the pre-storm epoch, but the burst hardness ratio was held constant at the average value. 
We then embedded these bursts randomly in time, adding simulated persistent emission where the hardness ratio varied with phase. 
We folded the hardness ratio curve using both burst photons and persistent emission within each burst’s T90 interval, and then calculated $\chi^2$ relative to a flat line, i.e., no phase dependence. 
Figure \ref{fig:burst_phase_distribution}d shows the $\chi^2$ distribution of the sampled bursts, which follows a $\chi^2$ distribution. 
The observed value deviates from the simulated samples by about 10$\sigma$, suggesting that the phase dependence is intrinsic to the bursts rather than caused by persistent emission. 
However, if we increase the persistent emission to roughly the flare peak level, i.e., about 100 times higher than the pre-storm level, the observed value deviates from the simulated samples at approximately the $4\sigma$ level. 
Therefore, we cannot securely draw the same conclusion for the burst-storm epoch.

\section{Discussion}\label{discussion}

The 2022 outburst from SGR 1935+2154 provided major insights into the physical conditions for FRB emission from magnetars. 
Timing analysis of the persistent emission has revealed two glitches bracketing the FRB 20221014A \citep{GiriAC2023, HuNE2024}. 
Between glitches, this source experienced a short-term outburst with a rapid flux increase approximately 2.5 hours before FRB 20221014A (two hours after the first glitch). 
This flux enhancement coincided with a burst storm, during which the burst occurrence rate increased from 0.6 bursts per minute to 7.2 bursts per minute. 
The burst storm ended with a flare-like event, although the total energy release was around $10^{41}$ erg, far lower than historical giant flares from sources such as SGR 0526$-$66, SGR 1806$-$20 and SGR 1900+14 \citep{EvansKL1980, HurleyCM1999, HurleyBS2005}. 
In this study, we define this event as an intermediate flare, which stands out as the most energetic burst in the 2022 activity. Yet this flare energy is still energetic enough that it may be detected at a few Mpc extragalactic distances (see below) by current X-ray instruments, suggesting similar contemporaneous but not directly simultaneous hard X-ray flares may present in FRBs in the local Universe.

High-cadence observations with \nustar\ enabled precise tracking of the timing and spectral evolution of this event. 
From the evolution of the hardness ratio, we identified a rapid softening in both the burst and persistent emission following the intermediate flare. 
This softening is supported by the following evidence from the detailed spectral analysis:
\begin{itemize}
    \item \textbf{Broadband spectral analysis of bursts}: \nicer\ and \nustar\ spectra of bright bursts detected before the intermediate flare show higher temperatures, with soft components ranging from $kT_1=1.5$--$2.1$ keV and hard components from $kT_2=4.4$--$6.8$ keV. 
    Post-flare bursts exhibit significantly lower temperatures ($kT_1=1.0$ keV and $kT_2=2.6$ keV), as shown in Figure \ref{fig:burst1904} and Table \ref{tab:bursrs_parameters}. 
    \item \textbf{Sudden hardness ratio drop}: The hardness ratio curve (Figure \ref{fig:burst_evolution_all}) revealed a sharp decrease in hardness ratio at the time of the intermediate flare.
    \item \textbf{Spectral evolution during the intermediate flare}: By dividing the intermediate flare into several epochs, we observed a rapid temperature drop within approximately one minute (see Figure \ref{fig:spectrum_huge_flare}). 
\end{itemize} 

These phenomenology suggest that the intermediate flare is associated with a qualitative step change in the activity of \src. This may result from an alteration of the magnetospheric configuration or the activation of a different region in the crust.

Further analysis of the hardness-intensity diagram reveals a counter-clockwise evolutionary pattern (Figure \ref{fig:persistent_hid}), with the intermediate flare positioned at the top, highlighting a significant shift in the spectral behavior of SGR 1935+2154. 
Contemporaneously, the burst occurrence rate gradually declined, as evidenced by the $\Delta t$ distribution. 
These altered conditions may be conducive to either FRB generation or escape, as the bright radio burst occurred soon thereafter.

The spectral behavior of X-ray burst (FRB-X) associated with FRB 20200428 can be described by a CPL of $\Gamma = 1.56 \pm 0.06$ and $E_f = 84^{+9}_{-7}$ keV \citep{LiLX2021}. 
The best-fit broadband X-ray fluence of this burst is $7.1 \times 10^{-7}$ erg cm$^{-2}$, corresponding to an energy release of $3.7 \times 10^{39}$ erg, assuming a distance of 6.6 kpc \citep{Zhou2020}. 
In comparison, the FRB 20221014A has a much lower fluence of $9.7 \pm 6.7$ kJy ms \citep{GiriAC2023}. 
The X-ray burst associated with FRB 20221014A has a reported fluence of $5.7\times10^{-7}$ erg cm$^{-2}$, implying a much higher X-ray-to-radio fluence ratio compared to that of FRB 20200428 \citep{RehanI2025}.
Nevertheless, these fluences are relatively low compared to giant flares, or even intermediate flares detected in this study, suggesting that FRBs are not directly connected to the most energetic flares in magnetars \citep{Wadiasingh2019, Wadiasingh2020, PearlmanSB2025}.

Using broadband \nicer\ and \emph{Fermi} GBM observations, \citet{YounesBK2021} showed that FRB-X exhibits a strikingly different spectral behavior compared to bursts in the burst storm. 
FRB-X displays a curved spectrum in the $E F_{E}$ plot, peaking at approximately 5 keV with a soft power-law tail. 
In contrast, burst spectra detected during the 2020 burst storm also have curved shapes but peak around 20 keV. 
Compared to FRB-X, the bursts in the 2020 burst storm have a harder photon index of $\Gamma \approx 0.5$ and lower e-folding energy of $E_f \lesssim 10$ keV. 
\citet{YounesBK2021} argued that FRB-X could originate from the polar region and may be highly collimated, which causes the spectral extension by the fireball encompassing higher altitudes.

The rapid spectral change preceded by the intermediate flare provides important insights into dynamic physical processes and conditions.
Although \nustar\ has a relatively narrow energy band compared to the combination of \nicer\ and \emph{Fermi} GBM, we observed similar spectral differences between bursts detected during the burst-storm and post-flare epochs (particularly post-flare 1 and 2). 
During the burst storm, the accumulated burst emission has a hard photon index of $\Gamma \approx 1.2$ and a low $E_f \approx 30$ keV, with the spectrum peaking at approximately 20--30 keV. 
In contrast, the accumulated burst emission observed in the post-flare epoch has a much softer $\Gamma \approx 2.3$, and $E_f$ cannot be well constrained within the \nustar\ energy band. 
While the spectral shape remains curved, the peak shifts to around 5 keV with a soft power-law tail. 
This is consistent with the result in the hardness-fluence diagram of bursts, of which the post-flare bursts systematically have much softer spectra compared to those detected in pre-storm and burst-storm epochs.
The rapid spectral evolution of the intermediate flare indicates an evolutionary timescale of 80s, suggesting relatively slow (compared to a neutron star light crossing timescale) changes in the super-Eddington conditions of the fireball, and its energy injection and evaporation. 
As short bursts by nature are confined fireballs whose spectral and luminosity properties depend on the active area and magnetic colatitude of footpoints samples, this evolution suggests an alteration or migration of the active zone of bursts in the crust.

The non-uniform phase dependence of bursts rates and hardness ratios (Figure~\ref{fig:burst_phase_distribution}) is not unique. In fact, a similar phenomenon was seen in \src\ by \citet{KanekoGB2021}. These observations confirms the picture that magnetar short bursts are confined fireballs with varied temperature along a flux tube, with hotter components \citep{IsraelRM2008,vanderHorst2012,YounesKv2014} associated with footpoints at low altitudes and cooler temperature components at high altitudes. The alignment of the hardness ratio of bursts with the persistent emission is likely geometric -- the observer lines of sight and spin phases where the hot burst footpoints are visible is also where the higher altitude non-thermal magnetospheric resonant Compton emission has field lines tangents toward the observer \citep[beamed from equatorial locales,][]{Wadiasingh2018}. 

This luminosity enhanced outburst emission, which attains $10^{37}$ erg/s (see Fig.~\ref{fig:persistent_hid}), must be partly magnetospheric. A neutrino thermostat limits crust thermal emission to $\sim10^{36}$ erg/s \citep{PonsR2012}, and cooling generally occurs on a timescale much longer than observed here. The luminosities and spectra here suggest optically thick conditions over extended zones of the magnetosphere, a possibly sustained confined and evaporating fireball prior to the ``clean'' lower luminosity conditions of the FRB. Note intermediate flare must also be an optically thick fireball, likely polar in locale spanning many stellar radii. This flare approaches $\sim 10^{40}$ erg/s, is clearly approaching or exceeding the magnetic Eddington limit of a magnetars, driving an outflow \citep[e.g.,][]{2016MNRAS.461..877V,Younes2023} and altering the magnetosphere of the magnetar, possibly dissipating preexisting twists. The strong stellar wind caused by the intermediate flare (and the burst storm) may carry away angular momentum from the system, significantly increasing the spin-down rate of the magnetar. 
This is consistent with the high spin-down rate observed between two glitches \citep{HuNE2024} or the apparent spin-down glitch with an uncertainty of 1--2 days \citep{Younes2023}.

These observations suggest that the intermediate flare clearly altered the output of \src\ for at least several hours, with the spectral behavior of bursts detected after the flare resembling that of FRB-X, suggesting a more polar origin to bursts.  
This challenges the idea that FRB-X-like bursts are rare, although the rarity of bright FRBs persists.

Finally, the intermediate flare has an average luminosity of about $4\times10^{39}$~erg~s$^{-1}$, likely underestimated due to \nustar's severe deadtime issues. 
Moreover, despite the high uncertainty due to the dead time issue, its peak luminosity exceeds the average by more than one order of magnitude, reaching $\gtrsim10^{41}$\,erg\,s$^{-1}$, which meets the definition of an intermediate flare \citep{TurollaZW2015}.
If such a flare occurred in a nearby galaxy, for instance M31, M81, or M82 \citep{MereghettiRS2024}, the corresponding fluxes could reach $\sim1\times10^{-10}$~erg~s$^{-1}$~cm$^{-2}$ to $\sim6\times10^{-12}$~erg~s$^{-1}$~cm$^{-2}$.
Both values are above the detection thresholds of non-imaging instruments (e.g., \nicer) and imaging telescopes (e.g., \emph{XMM-Newton}, \emph{Chandra}, and future X-ray missions such as AXIS or Athena).
These events would appear as fast X-ray transients \citep[see, e.g.,][]{DillmannMS2025, QuirolaBJ2024, BauerTS2017}, and combining such observations with radio monitoring programs like CHIME or BURSTT \citep{LinLL2022} could potentially uncover more examples linking magnetars and FRBs in nearby galaxies.

\section{Summary}
\label{summary}

The 2022 outburst of SGR 1935+2154 has offered valuable insights into the connection between magnetars and FRBs. 
During this activity, a short-term outburst occurred accompanied by a burst storm. 
At the end of the burst storm, an intermediate X-ray flare with an energy output of approximately $6.3\times10^{40}$ erg was observed, occurring 2.5 hours after the first spin-up glitch and 1.9 hours before FRB 20221014A. 
The evolution of the hardness ratio and detailed spectral analysis showed a significant softening of the emission within about a minute during the flare. 
During the burst storm, \src\ transitions from a high-hard state to a high-soft state, with the intermediate flare serving as a key turning point in the hardness–intensity diagram.
The time-resolved spectroscopy of this intermediate flare revealed a rapid spectral change. 
While similar spectral softening has been observed in a few bursts using Fermi-GBM data \citep{OzgeGK2024}, this is the first time where a flare has been observed to induce a permanent change in both the persistent and burst spectra. 
Most importantly, such a flare is energetic enough to be detected in a few Mpc, indicating that these events could potentially be observed in nearby galaxies.

We propose that the flare generated a strong wind that cleared the magnetosphere, untwisting the magnetic field lines and enabling radio emission from the quasi-polar regions. 
Although the narrow opening angle of the radio emission limits its detectability, X-ray bursts resembling FRB-X became more frequent after the flare, possibly indicating a cooling phase as the flare expanded. 
This behavior aligns with the characteristics of a fireball ejection from the magnetar. 
While FRB 20221014A is not directly tied to giant-flare-like events, such energetic flares appear to help create a suitable environment for radio emission for at least a few hours.

We further examined correlations between burst parameters, including $T_{90}$, $\Delta t$, fluence, hardness ratio, and pulse phase. 
The most intriguing finding is the hardness-fluence relationship, where bursts from the pre-storm, burst-storm, and post-flare epochs display a pattern resembling that of the total emission in the hardness-intensity diagram. 
This suggests that previous burst samples from other magnetars may exhibit biased correlations if they do not cover the full evolutionary stages. 
Additionally, the pulse phase distribution of burst hardness ratios reveals phase-dependent variations. 
The folded hardness ratio of burst photons aligns with that of the folded persistent emission, though the burst emission is slightly softer.
Time-resolved spectral analysis indicates that the persistent emission follows a similar evolutionary pattern as the burst emission. 
These suggest that the persistent emission and the burst emission come from similar active regions or at least are visible geometrically from similar locations.

The hardness-intensity shift accompanied by a change in burst waiting times suggests multiple active regions or different locales may be involved in state changes associated temporally with the intermediate flare. 
As short bursts can originate from Comptonized fireballs trapped in flux tubes, the spectral extent, flux and shape of bursts can depend on the involved flux tube. The evolution in the hardness-intensity diagram suggest at least three different locales involved in bursts.
This may result also from differing evolution of stresses in the crust where the short bursts are triggered. The strong phase dependence of the burst hardness suggests they originate from relatively low altitudes, with a fixed and localized active region. 

More intriguingly, a clear change in the burst waiting time statistics are established post-flare. This is compatible with a drastically different state of stress in the crust. The intermediate flare modified these stresses (likely relieved) in certain locales of the crust, modifying the system's ensuing dynamical evolution. 
Such a log-normal distribution of waiting times post-flare, reminiscent of that seen in some FRBs, is a signature of self-organized criticality. 

\src\ is currently the only known magnetar connected to FRBs. 
Additional samples of magnetar-FRB connections and potential bridging sources that fill the luminosity gap between FRBs from \src\ and the broader FRB population are needed. 
Future high-cadence, broadband monitoring of \src\ and other magnetar outbursts will be crucial for revealing unbiased correlations between these parameters.

\begin{acknowledgments}
We thank the anonymous referee for valuable comments that improved this paper. 
We also thank Dr. Brian Grefenstette, Principal Mission Scientist for \nustar, for his helpful comments in addressing the pile-up issue.
This work was supported by the National Aeronautics and Space Administration (NASA) through the \nicer\ mission and the Astrophysics Explorers Program. 
This research has also made use of data obtained with \nustar, a project led by Caltech, funded by NASA and managed by NASA/JPL, and has utilized the NUSTARDAS software package, jointly developed by the ASDC (Italy) and Caltech (USA). 
This research has made use of data and software provided by the High Energy Astrophysics Science Archive Research Center (HEASARC), which is a service of the Astrophysics Science Division at NASA/GSFC and the High Energy Astrophysics Division of the Smithsonian Astrophysical Observatory. 
C.-P.H. acknowledges support from the National Science and Technology Council in Taiwan through grant 112-2112-M-018-004-MY3.
W.C.G.H. acknowledges support through grant 80NSSC23K0078 from NASA.
Z.W. acknowledges support by NASA under award numbers 80GSFC21M0002 and 80GSFC24M0006. 
M.G.B. acknowledges support through grant 80NSSC22K0777 from NASA.
S.G. acknowledges the support of the CNES.
M.L.B. acknowledges support from the Max Planck Society.
R.S. is funded in part by NASA grants 80NSSC21K1997, 80NSSC23K1114, and 80NSSC25K7257 (PI G.Y.)
\end{acknowledgments}

\facilities{\emph{NICER}, \emph{NuSTAR}}
\software{HEASOFT}

\bibliographystyle{aasjournal}
\bibliography{reference}

\begin{thebibliography}{}
\expandafter\ifx\csname natexlab\endcsname\relax\def\natexlab#1{#1}\fi
\providecommand{\url}[1]{\href{#1}{#1}}
\providecommand{\dodoi}[1]{doi:~\href{http://doi.org/#1}{\nolinkurl{#1}}}
\providecommand{\doeprint}[1]{\href{http://ascl.net/#1}{\nolinkurl{http://ascl.net/#1}}}
\providecommand{\doarXiv}[1]{\href{https://arxiv.org/abs/#1}{\nolinkurl{https://arxiv.org/abs/#1}}}

\bibitem[{{Abac} {et~al.}(2024){Abac}, {Abbott}, {Abouelfettouh}, {Acernese},
  {Ackley}, {Adhicary}, {Adhikari}, {Adhikari}, {Adkins}, {Agarwal}, {Agathos},
  {Aghaei Abchouyeh}, {Aguiar}, {Aguilar}, {Aiello}, {Ain}, {Ajith}, {Akutsu},
  {Albanesi}, {Alfaidi}, {Al-Jodah}, {All{\'e}n{\'e}}, {Allocca},
  {Al-Shammari}, {Altin}, {Alvarez-Lopez}, {Amato}, {Amez-Droz}, {Amorosi},
  {Amra}, {Ananyeva}, {Anderson}, {Anderson}, {Andia}, {Ando}, {Andrade},
  {Andres}, {Andr{\'e}s-Carcasona}, {Andri{\'c}}, {Anglin}, {Ansoldi},
  {Antelis}, {Antier}, {Aoumi}, {Appavuravther}, {Appert}, {Apple}, {Arai},
  {Araya}, {Araya}, {Areeda}, {Argianas}, {Aritomi}, {Armato}, {Arnaud},
  {Arogeti}, {Aronson}, {Ashton}, {Aso}, {Assiduo}, {Assis de Souza Melo},
  {Aston}, {Astone}, {Attadio}, {Aubin}, {Aultoneal}, {Avallone}, {Azrad},
  {Babak}, {Badaracco}, {Badger}, {Bae}, {Bagnasco}, {Bagui}, {Baier},
  {Baiotti}, {Bajpai}, {Baka}, {Ball}, {Ballardin}, {Ballmer}, {Banagiri},
  {Banerjee}, {Bankar}, {Baral}, {Barayoga}, {Barish}, {Barker}, {Barneo},
  {Barone}, {Barr}, {Barsotti}, {Barsuglia}, {Barta}, {Bartoletti}, {Barton},
  {Bartos}, {Basak}, {Basalaev}, {Bassiri}, {Basti}, {Bates}, {Bawaj}, {Baxi},
  {Bayley}, {Baylor}, {Baynard}, {Bazzan}, {Bedakihale}, {Beirnaert}, {Bejger},
  {Belardinelli}, {Bell}, {Benedetto}, {Benoit}, {Bentley}, {Ben Yaala},
  {Bera}, {Berbel}, {Bergamin}, {Berger}, {Bernuzzi}, {Beroiz}, {Bersanetti},
  {Bertolini}, {Betzwieser}, {Beveridge}, {Bevins}, {Bhandare}, {Bhardwaj},
  {Bhatt}, {Bhattacharjee}, {Bhaumik}, {Bhowmick}, {Bianchi}, {Bilenko},
  {Billingsley}, {Binetti}, {Bini}, {Birnholtz}, {Biscoveanu}, {Bisht},
  {Bitossi}, {Bizouard}, {Blackburn}, {Blagg}, {Blair}, {Blair}, {Bobba},
  {Bode}, {Boileau}, {Boldrini}, {Bolingbroke}, {Bolliand}, {Bonavena},
  {Bondarescu}, {Bondu}, {Bonilla}, {Bonilla}, {Bonino}, {Bonnand}, {Booker},
  {Borchers}, {Boschi}, {Bose}, {Bossilkov}, {Boudart}, {Boudon}, {Bozzi},
  {Bradaschia}, {Brady}, {Braglia}, {Branch}, {Branchesi}, {Brandt}, {Braun},
  {Breschi}, {Briant}, {Brillet}, {Brinkmann}, {Brockill}, {Brockmueller},
  {Brooks}, {Brown}, {Brown}, {Brozzetti}, {Brunett}, {Bruno}, {Bruntz},
  {Bryant}, {Bucci}, {Buchanan}, {Bulashenko}, {Bulik}, {Bulten}, {Buonanno},
  {Burtnyk}, {Buscicchio}, {Buskulic}, \& {Buy}}]{LVK2024_SGR1935}
{Abac}, A.~G., {Abbott}, R., {Abouelfettouh}, I., {et~al.} 2024, \apj, 977,
  255, \dodoi{10.3847/1538-4357/ad8de0}

\bibitem[{{Bachetti} {et~al.}(2021){Bachetti}, {Markwardt}, {Grefenstette},
  {Gotthelf}, {Kuiper}, {Barret}, {Cook}, {Davis}, {F{\"u}rst}, {Forster},
  {Harrison}, {Madsen}, {Miyasaka}, {Roberts}, {Tomsick}, \&
  {Walton}}]{BachettiMG2021}
{Bachetti}, M., {Markwardt}, C.~B., {Grefenstette}, B.~W., {et~al.} 2021, \apj,
  908, 184, \dodoi{10.3847/1538-4357/abd1d6}

\bibitem[{{Bailes}(2022)}]{Bailes2022}
{Bailes}, M. 2022, Science, 378, abj3043, \dodoi{10.1126/science.abj3043}

\bibitem[{{Bauer} {et~al.}(2017){Bauer}, {Treister}, {Schawinski}, {Schulze},
  {Luo}, {Alexander}, {Brandt}, {Comastri}, {Forster}, {Gilli}, {Kann},
  {Maeda}, {Nomoto}, {Paolillo}, {Ranalli}, {Schneider}, {Shemmer}, {Tanaka},
  {Tolstov}, {Tominaga}, {Tozzi}, {Vignali}, {Wang}, {Xue}, \&
  {Yang}}]{BauerTS2017}
{Bauer}, F.~E., {Treister}, E., {Schawinski}, K., {et~al.} 2017, \mnras, 467,
  4841, \dodoi{10.1093/mnras/stx417}

\bibitem[{{Beloborodov}(2017)}]{Beloborodov2017}
{Beloborodov}, A.~M. 2017, \apjl, 843, L26, \dodoi{10.3847/2041-8213/aa78f3}

\bibitem[{{Beniamini} {et~al.}(2025){Beniamini}, {Wadiasingh}, {Trigg},
  {Chirenti}, {Burns}, {Younes}, {Negro}, \& {Granot}}]{BeniaminiWT2024}
{Beniamini}, P., {Wadiasingh}, Z., {Trigg}, A., {et~al.} 2025, \apj, 980, 211,
  \dodoi{10.3847/1538-4357/ada947}

\bibitem[{{Bochenek} {et~al.}(2020){Bochenek}, {Ravi}, {Belov}, {Hallinan},
  {Kocz}, {Kulkarni}, \& {McKenna}}]{BochenekRB2020}
{Bochenek}, C.~D., {Ravi}, V., {Belov}, K.~V., {et~al.} 2020, \nat, 587, 59,
  \dodoi{10.1038/s41586-020-2872-x}

\bibitem[{{Burns} {et~al.}(2021){Burns}, {Svinkin}, {Hurley}, {Wadiasingh},
  {Negro}, {Younes}, {Hamburg}, {Ridnaia}, {Cook}, {Cenko}, {Aloisi}, {Ashton},
  {Baring}, {Briggs}, {Christensen}, {Frederiks}, {Goldstein}, {Hui}, {Kaplan},
  {Kasliwal}, {Kocevski}, {Roberts}, {Savchenko}, {Tohuvavohu}, {Veres}, \&
  {Wilson-Hodge}}]{BurnsSH2021}
{Burns}, E., {Svinkin}, D., {Hurley}, K., {et~al.} 2021, \apjl, 907, L28,
  \dodoi{10.3847/2041-8213/abd8c8}

\bibitem[{{CHIME/FRB Collaboration} {et~al.}(2020){CHIME/FRB Collaboration},
  {Andersen}, {Bandura}, {Bhardwaj}, {Bij}, {Boyce}, {Boyle}, {Brar},
  {Cassanelli}, {Chawla}, {Chen}, {Cliche}, {Cook}, {Cubranic}, {Curtin},
  {Denman}, {Dobbs}, {Dong}, {Fandino}, {Fonseca}, {Gaensler}, {Giri}, {Good},
  {Halpern}, {Hill}, {Hinshaw}, {H{\"o}fer}, {Josephy}, {Kania}, {Kaspi},
  {Landecker}, {Leung}, {Li}, {Lin}, {Masui}, {McKinven}, {Mena-Parra},
  {Merryfield}, {Meyers}, {Michilli}, {Milutinovic}, {Mirhosseini},
  {M{\"u}nchmeyer}, {Naidu}, {Newburgh}, {Ng}, {Patel}, {Pen},
  {Pinsonneault-Marotte}, {Pleunis}, {Quine}, {Rafiei-Ravandi}, {Rahman},
  {Ransom}, {Renard}, {Sanghavi}, {Scholz}, {Shaw}, {Shin}, {Siegel}, {Singh},
  {Smegal}, {Smith}, {Stairs}, {Tan}, {Tendulkar}, {Tretyakov}, {Vanderlinde},
  {Wang}, {Wulf}, \& {Zwaniga}}]{CHIME2020_SGR1935}
{CHIME/FRB Collaboration}, {Andersen}, B.~C., {Bandura}, K.~M., {et~al.} 2020,
  \nat, 587, 54, \dodoi{10.1038/s41586-020-2863-y}

\bibitem[{{CHIME/FRB Collaboration} {et~al.}(2021){CHIME/FRB Collaboration},
  {Amiri}, {Andersen}, {Bandura}, {Berger}, {Bhardwaj}, {Boyce}, {Boyle},
  {Brar}, {Breitman}, {Cassanelli}, {Chawla}, {Chen}, {Cliche}, {Cook},
  {Cubranic}, {Curtin}, {Deng}, {Dobbs}, {Dong}, {Eadie}, {Fandino}, {Fonseca},
  {Gaensler}, {Giri}, {Good}, {Halpern}, {Hill}, {Hinshaw}, {Josephy},
  {Kaczmarek}, {Kader}, {Kania}, {Kaspi}, {Landecker}, {Lang}, {Leung}, {Li},
  {Lin}, {Masui}, {McKinven}, {Mena-Parra}, {Merryfield}, {Meyers}, {Michilli},
  {Milutinovic}, {Mirhosseini}, {M{\"u}nchmeyer}, {Naidu}, {Newburgh}, {Ng},
  {Patel}, {Pen}, {Petroff}, {Pinsonneault-Marotte}, {Pleunis},
  {Rafiei-Ravandi}, {Rahman}, {Ransom}, {Renard}, {Sanghavi}, {Scholz}, {Shaw},
  {Shin}, {Siegel}, {Sikora}, {Singh}, {Smith}, {Stairs}, {Tan}, {Tendulkar},
  {Vanderlinde}, {Wang}, {Wulf}, \& {Zwaniga}}]{CHIME_Catalog_2021}
{CHIME/FRB Collaboration}, {Amiri}, M., {Andersen}, B.~C., {et~al.} 2021,
  \apjs, 257, 59, \dodoi{10.3847/1538-4365/ac33ab}

\bibitem[{{Dillmann} {et~al.}(2025){Dillmann}, {Mart{\'\i}nez-Galarza},
  {Soria}, {Stefano}, \& {Kashyap}}]{DillmannMS2025}
{Dillmann}, S., {Mart{\'\i}nez-Galarza}, J.~R., {Soria}, R., {Stefano}, R.~D.,
  \& {Kashyap}, V.~L. 2025, \mnras, 537, 931, \dodoi{10.1093/mnras/stae2808}

\bibitem[{{Dong} \& {Chime/Frb Collaboration}(2022)}]{Dong_CHIME_2022}
{Dong}, F.~A., \& {Chime/Frb Collaboration}. 2022, The Astronomer's Telegram,
  15681, 1

\bibitem[{{Duncan} \& {Thompson}(1992)}]{DuncanT1992}
{Duncan}, R.~C., \& {Thompson}, C. 1992, \apjl, 392, L9, \dodoi{10.1086/186413}

\bibitem[{{Enoto} {et~al.}(2019){Enoto}, {Kisaka}, \& {Shibata}}]{EnotoKS2019}
{Enoto}, T., {Kisaka}, S., \& {Shibata}, S. 2019, Reports on Progress in
  Physics, 82, 106901, \dodoi{10.1088/1361-6633/ab3def}

\bibitem[{{Evans} {et~al.}(1980){Evans}, {Klebesadel}, {Laros}, {Cline},
  {Desai}, {Teegarden}, {Pizzichini}, {Hurley}, {Niel}, \&
  {Vedrenne}}]{EvansKL1980}
{Evans}, W.~D., {Klebesadel}, R.~W., {Laros}, J.~G., {et~al.} 1980, \apjl, 237,
  L7, \dodoi{10.1086/183222}

\bibitem[{{Gavriil} {et~al.}(2004){Gavriil}, {Kaspi}, \& {Woods}}]{Gavriil2004}
{Gavriil}, F.~P., {Kaspi}, V.~M., \& {Woods}, P.~M. 2004, \apj, 607, 959,
  \dodoi{10.1086/383564}

\bibitem[{{Gendreau} {et~al.}(2016){Gendreau}, {Arzoumanian}, {Adkins},
  {Albert}, {Anders}, {Aylward}, {Baker}, {Balsamo}, {Bamford}, {Benegalrao},
  {Berry}, {Bhalwani}, {Black}, {Blaurock}, {Bronke}, {Brown}, {Budinoff},
  {Cantwell}, {Cazeau}, {Chen}, {Clement}, {Colangelo}, {Coleman},
  {Coopersmith}, {Dehaven}, {Doty}, {Egan}, {Enoto}, {Fan}, {Ferro}, {Foster},
  {Galassi}, {Gallo}, {Green}, {Grosh}, {Ha}, {Hasouneh}, {Heefner}, {Hestnes},
  {Hoge}, {Jacobs}, {J{\o}rgensen}, {Kaiser}, {Kellogg}, {Kenyon}, {Koenecke},
  {Kozon}, {LaMarr}, {Lambertson}, {Larson}, {Lentine}, {Lewis}, {Lilly},
  {Liu}, {Malonis}, {Manthripragada}, {Markwardt}, {Matonak}, {Mcginnis},
  {Miller}, {Mitchell}, {Mitchell}, {Mohammed}, {Monroe}, {Montt de Garcia},
  {Mul{\'e}}, {Nagao}, {Ngo}, {Norris}, {Norwood}, {Novotka}, {Okajima},
  {Olsen}, {Onyeachu}, {Orosco}, {Peterson}, {Pevear}, {Pham}, {Pollard},
  {Pope}, {Powers}, {Powers}, {Price}, {Prigozhin}, {Ramirez}, {Reid},
  {Remillard}, {Rogstad}, {Rosecrans}, {Rowe}, {Sager}, {Sanders}, {Savadkin},
  {Saylor}, {Schaeffer}, {Schweiss}, {Semper}, {Serlemitsos}, {Shackelford},
  {Soong}, {Struebel}, {Vezie}, {Villasenor}, {Winternitz}, {Wofford},
  {Wright}, {Yang}, \& {Yu}}]{GendreauAA2016}
{Gendreau}, K.~C., {Arzoumanian}, Z., {Adkins}, P.~W., {et~al.} 2016, in
  Society of Photo-Optical Instrumentation Engineers (SPIE) Conference Series,
  Vol. 9905, Space Telescopes and Instrumentation 2016: Ultraviolet to Gamma
  Ray, ed. J.-W.~A. {den Herder}, T.~{Takahashi}, \& M.~{Bautz}, 99051H,
  \dodoi{10.1117/12.2231304}

\bibitem[{{Giri} {et~al.}(2023){Giri}, {Andersen}, {Chawla}, {Curtin},
  {Fonseca}, {Kaspi}, {Lin}, {Masui}, {Sand}, {Scholz}, {Abbott}, {Dong},
  {Gaensler}, {Leung}, {Michilli}, {Bhardwaj}, {M{\"u}nchmeyer}, {Pandhi},
  {Pearlman}, {Pleunis}, {Rafiei-Ravandi}, {Reda}, {Shin}, {Smith}, {Stairs},
  {Stenning}, \& {Tendulkar}}]{GiriAC2023}
{Giri}, U., {Andersen}, B.~C., {Chawla}, P., {et~al.} 2023, arXiv e-prints,
  arXiv:2310.16932, \dodoi{10.48550/arXiv.2310.16932}

\bibitem[{{G{\"o}{\v{g}}{\"u}{\c{S}} } {et~al.}(1999){G{\"o}{\v{g}}{\"u}{\c{S}}
  }, {Woods}, {Kouveliotou}, {van Paradijs}, {Briggs}, {Duncan}, \&
  {Thompson}}]{GogusWK1999}
{G{\"o}{\v{g}}{\"u}{\c{S}} }, E., {Woods}, P.~M., {Kouveliotou}, C., {et~al.}
  1999, \apjl, 526, L93, \dodoi{10.1086/312380}

\bibitem[{{G{\"o}{\v{g}}{\"u}{\c{s}}}
  {et~al.}(2001){G{\"o}{\v{g}}{\"u}{\c{s}}}, {Kouveliotou}, {Woods},
  {Thompson}, {Duncan}, \& {Briggs}}]{GogusKW2001}
{G{\"o}{\v{g}}{\"u}{\c{s}}}, E., {Kouveliotou}, C., {Woods}, P.~M., {et~al.}
  2001, \apj, 558, 228, \dodoi{10.1086/322463}

\bibitem[{{G{\"o}{\v{g}}{\"u}{\c{s}}}
  {et~al.}(2000){G{\"o}{\v{g}}{\"u}{\c{s}}}, {Woods}, {Kouveliotou}, {van
  Paradijs}, {Briggs}, {Duncan}, \& {Thompson}}]{GogusWK2000}
{G{\"o}{\v{g}}{\"u}{\c{s}}}, E., {Woods}, P.~M., {Kouveliotou}, C., {et~al.}
  2000, \apjl, 532, L121, \dodoi{10.1086/312583}

\bibitem[{{Grefenstette} {et~al.}(2016){Grefenstette}, {Glesener}, {Krucker},
  {Hudson}, {Hannah}, {Smith}, {Vogel}, {White}, {Madsen}, {Marsh}, {Caspi},
  {Chen}, {Shih}, {Kuhar}, {Boggs}, {Christensen}, {Craig}, {Forster},
  {Hailey}, {Harrison}, {Miyasaka}, {Stern}, \& {Zhang}}]{GrefenstetteGK2016}
{Grefenstette}, B.~W., {Glesener}, L., {Krucker}, S., {et~al.} 2016, \apj, 826,
  20, \dodoi{10.3847/0004-637X/826/1/20}

\bibitem[{{Hu} {et~al.}(2024){Hu}, {Narita}, {Enoto}, {Younes}, {Wadiasingh},
  {Baring}, {Ho}, {Guillot}, {Ray}, {G{\"u}ver}, {Rajwade}, {Arzoumanian},
  {Kouveliotou}, {Harding}, \& {Gendreau}}]{HuNE2024}
{Hu}, C.-P., {Narita}, T., {Enoto}, T., {et~al.} 2024, \nat, 626, 500,
  \dodoi{10.1038/s41586-023-07012-5}

\bibitem[{{Hurley} {et~al.}(1999){Hurley}, {Cline}, {Mazets}, {Barthelmy},
  {Butterworth}, {Marshall}, {Palmer}, {Aptekar}, {Golenetskii}, {Il'Inskii},
  {Frederiks}, {McTiernan}, {Gold}, \& {Trombka}}]{HurleyCM1999}
{Hurley}, K., {Cline}, T., {Mazets}, E., {et~al.} 1999, \nat, 397, 41,
  \dodoi{10.1038/16199}

\bibitem[{{Hurley} {et~al.}(2005){Hurley}, {Boggs}, {Smith}, {Duncan}, {Lin},
  {Zoglauer}, {Krucker}, {Hurford}, {Hudson}, {Wigger}, {Hajdas}, {Thompson},
  {Mitrofanov}, {Sanin}, {Boynton}, {Fellows}, {von Kienlin}, {Lichti}, {Rau},
  \& {Cline}}]{HurleyBS2005}
{Hurley}, K., {Boggs}, S.~E., {Smith}, D.~M., {et~al.} 2005, \nat, 434, 1098,
  \dodoi{10.1038/nature03519}

\bibitem[{{Ibrahim} {et~al.}(2024){Ibrahim}, {Borghese}, {Coti Zelati},
  {Parent}, {Marino}, {Ould-Boukattine}, {Rea}, {Ascenzi}, {Pacholski},
  {Mereghetti}, {Israel}, {Tiengo}, {Possenti}, {Burgay}, {Turolla}, {Zane},
  {Esposito}, {G{\"o}tz}, {Campana}, {Kirsten}, {Gawro{\'n}ski}, \&
  {Hessels}}]{IbrahimBC2024}
{Ibrahim}, A.~Y., {Borghese}, A., {Coti Zelati}, F., {et~al.} 2024, \apj, 965,
  87, \dodoi{10.3847/1538-4357/ad293b}

\bibitem[{{Israel} {et~al.}(2016){Israel}, {Esposito}, {Rodr{\'\i}guez
  Castillo}, \& {Sidoli}}]{Israel2016}
{Israel}, G.~L., {Esposito}, P., {Rodr{\'\i}guez Castillo}, G.~A., \& {Sidoli},
  L. 2016, \mnras, 462, 4371, \dodoi{10.1093/mnras/stw1897}

\bibitem[{{Israel} {et~al.}(2008){Israel}, {Romano}, {Mangano}, {Dall'Osso},
  {Chincarini}, {Stella}, {Campana}, {Belloni}, {Tagliaferri}, {Blustin},
  {Sakamoto}, {Hurley}, {Zane}, {Moretti}, {Palmer}, {Guidorzi}, {Burrows},
  {Gehrels}, \& {Krimm}}]{IsraelRM2008}
{Israel}, G.~L., {Romano}, P., {Mangano}, V., {et~al.} 2008, \apj, 685, 1114,
  \dodoi{10.1086/590486}

\bibitem[{{Kaneko} {et~al.}(2021){Kaneko}, {G{\"o}{\u{g}}{\"u}{\c{s}}},
  {Baring}, {Kouveliotou}, {Lin}, {Roberts}, {van der Horst}, {Younes},
  {Keskin}, \& {{\c{C}}oban}}]{KanekoGB2021}
{Kaneko}, Y., {G{\"o}{\u{g}}{\"u}{\c{s}}}, E., {Baring}, M.~G., {et~al.} 2021,
  \apjl, 916, L7, \dodoi{10.3847/2041-8213/ac0fe7}

\bibitem[{{Kaspi} \& {Beloborodov}(2017)}]{KaspiB2017}
{Kaspi}, V.~M., \& {Beloborodov}, A.~M. 2017, \araa, 55, 261,
  \dodoi{10.1146/annurev-astro-081915-023329}

\bibitem[{{Keane} {et~al.}(2012){Keane}, {Stappers}, {Kramer}, \&
  {Lyne}}]{KeaneSK2012}
{Keane}, E.~F., {Stappers}, B.~W., {Kramer}, M., \& {Lyne}, A.~G. 2012, \mnras,
  425, L71, \dodoi{10.1111/j.1745-3933.2012.01306.x}

\bibitem[{{Keskin} {et~al.}(2024){Keskin}, {G{\"o}{\u{g}}{\"u}{\c{s}}},
  {Kaneko}, {Demirer}, {Yamasaki}, {Baring}, {Lin}, {Roberts}, \&
  {Kouveliotou}}]{OzgeGK2024}
{Keskin}, {\"O}., {G{\"o}{\u{g}}{\"u}{\c{s}}}, E., {Kaneko}, Y., {et~al.} 2024,
  \apj, 965, 130, \dodoi{10.3847/1538-4357/ad2fce}

\bibitem[{{Kirsten} {et~al.}(2021){Kirsten}, {Snelders}, {Jenkins}, {Nimmo},
  {van den Eijnden}, {Hessels}, {Gawro{\'n}ski}, \& {Yang}}]{Kirsten2020}
{Kirsten}, F., {Snelders}, M.~P., {Jenkins}, M., {et~al.} 2021, Nature
  Astronomy, 5, 414, \dodoi{10.1038/s41550-020-01246-3}

\bibitem[{{Kouveliotou} {et~al.}(1993){Kouveliotou}, {Meegan}, {Fishman},
  {Bhat}, {Briggs}, {Koshut}, {Paciesas}, \& {Pendleton}}]{KouveliotouMF1993}
{Kouveliotou}, C., {Meegan}, C.~A., {Fishman}, G.~J., {et~al.} 1993, \apjl,
  413, L101, \dodoi{10.1086/186969}

\bibitem[{{Kumar} {et~al.}(2017){Kumar}, {Lu}, \& {Bhattacharya}}]{Kumar2017}
{Kumar}, P., {Lu}, W., \& {Bhattacharya}, M. 2017, \mnras, 468, 2726,
  \dodoi{10.1093/mnras/stx665}

\bibitem[{{Li} {et~al.}(2021){Li}, {Lin}, {Xiong}, {Ge}, {Li}, {Li}, {Lu},
  {Zhang}, {Tuo}, {Nang}, {Zhang}, {Xiao}, {Chen}, {Song}, {Xu}, {Liu}, {Jia},
  {Cao}, {Qu}, {Zhang}, {Gu}, {Liao}, {Zhao}, {Tan}, {Nie}, {Zhao}, {Zheng},
  {Zheng}, {Luo}, {Cai}, {Li}, {Xue}, {Bu}, {Chang}, {Chen}, {Chen}, {Chen},
  {Chen}, {Chen}, {Cui}, {Cui}, {Deng}, {Dong}, {Du}, {Fu}, {Gao}, {Gao},
  {Gao}, {Gu}, {Guan}, {Guo}, {Han}, {Huang}, {Huo}, {Jiang}, {Jiang}, {Jin},
  {Jin}, {Kong}, {Li}, {Li}, {Li}, {Li}, {Li}, {Li}, {Li}, {Liang}, {Liu},
  {Liu}, {Liu}, {Liu}, {Liu}, {Lu}, {Lu}, {Luo}, {Ma}, {Meng}, {Ou}, {Sai},
  {Shang}, {Song}, {Sun}, {Tao}, {Wang}, {Wang}, {Wang}, {Wang}, {Wang}, {Wen},
  {Wu}, {Wu}, {Wu}, {Xiao}, {Xu}, {Yang}, {Yang}, {Yang}, {Yang}, {Yi}, {Yin},
  {You}, {Zhang}, {Zhang}, {Zhang}, {Zhang}, {Zhang}, {Zhang}, {Zhang},
  {Zhang}, {Zhang}, {Zhang}, {Zhang}, {Zhang}, {Zhang}, {Zhang}, {Zhang},
  {Zhang}, {Zhou}, {Zhou}, {Zhu}, {Zhu}, \& {Zhuang}}]{LiLX2021}
{Li}, C.~K., {Lin}, L., {Xiong}, S.~L., {et~al.} 2021, Nature Astronomy, 5,
  378, \dodoi{10.1038/s41550-021-01302-6}

\bibitem[{{Lin} {et~al.}(2022){Lin}, {Lin}, {Li}, {Tseng}, {Jiang}, {Wang},
  {Cheng}, {Pen}, {Chen}, {Chen}, {Chen}, {Goto}, {Hashimoto}, {Hwang}, {King},
  {Kubo}, {Kuo}, {Mills}, {Nam}, {Oshiro}, {Shen}, {Tseng}, {Wang}, {Wu},
  {Bower}, {Chang}, {Chen}, {Chen}, {Chiang}, {Fedynitch}, {Gusinskaia}, {Ho},
  {Hsiao}, {Hu}, {Huang}, {J{\'a}uregui Garc{\'\i}a}, {Kim}, {Kuo}, {Ling},
  {On}, {Peterson}, {R. Raquel}, {Su}, {Uno}, {Wu}, {Yamasaki}, \&
  {Zhu}}]{LinLL2022}
{Lin}, H.-H., {Lin}, K.-y., {Li}, C.-T., {et~al.} 2022, \pasp, 134, 094106,
  \dodoi{10.1088/1538-3873/ac8f71}

\bibitem[{{Linscott} \& {Erkes}(1980)}]{Linscott1980}
{Linscott}, I.~R., \& {Erkes}, J.~W. 1980, \apjl, 236, L109,
  \dodoi{10.1086/183209}

\bibitem[{{Lorimer} {et~al.}(2007){Lorimer}, {Bailes}, {McLaughlin},
  {Narkevic}, \& {Crawford}}]{LorimerBM2007}
{Lorimer}, D.~R., {Bailes}, M., {McLaughlin}, M.~A., {Narkevic}, D.~J., \&
  {Crawford}, F. 2007, Science, 318, 777, \dodoi{10.1126/science.1147532}

\bibitem[{{Lyubarsky}(2014)}]{Lyubarsky2014}
{Lyubarsky}, Y. 2014, \mnras, 442, L9, \dodoi{10.1093/mnrasl/slu046}

\bibitem[{{Lyutikov}(2017)}]{Lyutikov2017}
{Lyutikov}, M. 2017, \apjl, 838, L13, \dodoi{10.3847/2041-8213/aa62fa}

\bibitem[{{Lyutikov}(2021)}]{Lyutikov2021}
---. 2021, \apj, 922, 166, \dodoi{10.3847/1538-4357/ac1b32}

\bibitem[{{Lyutikov} \& {Popov}(2020{\natexlab{a}})}]{Lyutikov2020}
{Lyutikov}, M., \& {Popov}, S. 2020{\natexlab{a}}, arXiv e-prints,
  arXiv:2005.05093.
\newblock \doarXiv{2005.05093}

\bibitem[{{Lyutikov} \& {Popov}(2020{\natexlab{b}})}]{Zhou2020}
---. 2020{\natexlab{b}}, arXiv e-prints, arXiv:2005.05093,
  \dodoi{10.48550/arXiv.2005.05093}

\bibitem[{{Maan} {et~al.}(2022){Maan}, {Leeuwen}, {Straal}, \&
  {Pastor-Marazuela}}]{MaanLS2022}
{Maan}, Y., {Leeuwen}, J.~v., {Straal}, S., \& {Pastor-Marazuela}, I. 2022, The
  Astronomer's Telegram, 15697, 1

\bibitem[{{Mazets} {et~al.}(1979){Mazets}, {Golenetskij}, \&
  {Guryan}}]{MazetsGG1979}
{Mazets}, E.~P., {Golenetskij}, S.~V., \& {Guryan}, Y.~A. 1979, Soviet
  Astronomy Letters, 5, 641

\bibitem[{{Mereghetti} {et~al.}(2022){Mereghetti}, {Gotz}, {Ferrigno}, {Bozzo},
  {Savchenko}, {Ducci}, \& {Borkowski}}]{MereghettiGF2022}
{Mereghetti}, S., {Gotz}, D., {Ferrigno}, C., {et~al.} 2022, GRB Coordinates
  Network, 32675, 1

\bibitem[{{Mereghetti} {et~al.}(2020){Mereghetti}, {Savchenko}, {Ferrigno},
  {G{\"o}tz}, {Rigoselli}, {Tiengo}, {Bazzano}, {Bozzo}, {Coleiro},
  {Courvoisier}, {Doyle}, {Goldwurm}, {Hanlon}, {Jourdain}, {von Kienlin},
  {Lutovinov}, {Martin-Carrillo}, {Molkov}, {Natalucci}, {Onori}, {Panessa},
  {Rodi}, {Rodriguez}, {S{\'a}nchez-Fern{\'a}ndez}, {Sunyaev}, \&
  {Ubertini}}]{MereghettiSF2020}
{Mereghetti}, S., {Savchenko}, V., {Ferrigno}, C., {et~al.} 2020, \apjl, 898,
  L29, \dodoi{10.3847/2041-8213/aba2cf}

\bibitem[{{Mereghetti} {et~al.}(2024){Mereghetti}, {Rigoselli}, {Salvaterra},
  {Pacholski}, {Rodi}, {Gotz}, {Arrigoni}, {D'Avanzo}, {Adami}, {Bazzano},
  {Bozzo}, {Brivio}, {Campana}, {Cappellaro}, {Chenevez}, {De Luise}, {Ducci},
  {Esposito}, {Ferrigno}, {Ferro}, {Israel}, {Le Floc'h}, {Martin-Carrillo},
  {Onori}, {Rea}, {Reguitti}, {Savchenko}, {Souami}, {Tartaglia}, {Thuillot},
  {Tiengo}, {Tomasella}, {Topinka}, {Turpin}, \& {Ubertini}}]{MereghettiRS2024}
{Mereghetti}, S., {Rigoselli}, M., {Salvaterra}, R., {et~al.} 2024, \nat, 629,
  58, \dodoi{10.1038/s41586-024-07285-4}

\bibitem[{{Metzger} {et~al.}(2019){Metzger}, {Margalit}, \&
  {Sironi}}]{Metzger2019}
{Metzger}, B.~D., {Margalit}, B., \& {Sironi}, L. 2019, \mnras, 485, 4091,
  \dodoi{10.1093/mnras/stz700}

\bibitem[{{Murase} {et~al.}(2016){Murase}, {Kashiyama}, \&
  {M{\'e}sz{\'a}ros}}]{MuraseKM2016}
{Murase}, K., {Kashiyama}, K., \& {M{\'e}sz{\'a}ros}, P. 2016, \mnras, 461,
  1498, \dodoi{10.1093/mnras/stw1328}

\bibitem[{{Paczynski}(1992)}]{Paczynski1992}
{Paczynski}, B. 1992, \actaa, 42, 145

\bibitem[{{Pearlman} {et~al.}(2025){Pearlman}, {Scholz}, {Bethapudi},
  {Hessels}, {Kaspi}, {Kirsten}, {Nimmo}, {Spitler}, {Fonseca}, {Meyers},
  {Stairs}, {Tan}, {Bhardwaj}, {Chatterjee}, {Cook}, {Curtin}, {Dong},
  {Eftekhari}, {Gaensler}, {G{\"u}ver}, {Kaczmarek}, {Leung}, {Masui},
  {Michilli}, {Prince}, {Sand}, {Shin}, {Smith}, \&
  {Tendulkar}}]{PearlmanSB2025}
{Pearlman}, A.~B., {Scholz}, P., {Bethapudi}, S., {et~al.} 2025, Nature
  Astronomy, 9, 111, \dodoi{10.1038/s41550-024-02386-6}

\bibitem[{{Petroff} {et~al.}(2019){Petroff}, {Hessels}, \&
  {Lorimer}}]{Petroff2019}
{Petroff}, E., {Hessels}, J.~W.~T., \& {Lorimer}, D.~R. 2019, \aapr, 27, 4,
  \dodoi{10.1007/s00159-019-0116-6}

\bibitem[{{Petroff} {et~al.}(2022){Petroff}, {Hessels}, \&
  {Lorimer}}]{PetroffHL2022}
---. 2022, \aapr, 30, 2, \dodoi{10.1007/s00159-022-00139-w}

\bibitem[{{Platts} {et~al.}(2019){Platts}, {Weltman}, {Walters}, {Tendulkar},
  {Gordin}, \& {Kandhai}}]{Platts2019}
{Platts}, E., {Weltman}, A., {Walters}, A., {et~al.} 2019, \physrep, 821, 1,
  \dodoi{10.1016/j.physrep.2019.06.003}

\bibitem[{{Pons} \& {Rea}(2012)}]{PonsR2012}
{Pons}, J.~A., \& {Rea}, N. 2012, \apjl, 750, L6,
  \dodoi{10.1088/2041-8205/750/1/L6}

\bibitem[{{Popov} \& {Postnov}(2010)}]{PopovP2010}
{Popov}, S.~B., \& {Postnov}, K.~A. 2010, in Evolution of Cosmic Objects
  through their Physical Activity, ed. H.~A. {Harutyunian}, A.~M. {Mickaelian},
  \& Y.~{Terzian}, 129--132, \dodoi{10.48550/arXiv.0710.2006}

\bibitem[{{Quirola-V{\'a}squez} {et~al.}(2024){Quirola-V{\'a}squez}, {Bauer},
  {Jonker}, {Brandt}, {Eappachen}, {Levan}, {L{\'o}pez}, {Luo}, {Ravasio},
  {Sun}, {Xue}, {Yang}, \& {Zheng}}]{QuirolaBJ2024}
{Quirola-V{\'a}squez}, J., {Bauer}, F.~E., {Jonker}, P.~G., {et~al.} 2024,
  \aap, 683, A243, \dodoi{10.1051/0004-6361/202347629}

\bibitem[{{Rehan} \& {Ibrahim}(2025)}]{RehanI2025}
{Rehan}, N.~S., \& {Ibrahim}, A.~I. 2025, \apjs, 276, 60,
  \dodoi{10.3847/1538-4365/ad95f9}

\bibitem[{{Ridnaia} {et~al.}(2021){Ridnaia}, {Svinkin}, {Frederiks}, {Bykov},
  {Popov}, {Aptekar}, {Golenetskii}, {Lysenko}, {Tsvetkova}, {Ulanov}, \&
  {Cline}}]{RidnaiaSF2021}
{Ridnaia}, A., {Svinkin}, D., {Frederiks}, D., {et~al.} 2021, Nature Astronomy,
  5, 372, \dodoi{10.1038/s41550-020-01265-0}

\bibitem[{{Scargle} {et~al.}(2013){Scargle}, {Norris}, {Jackson}, \&
  {Chiang}}]{ScargleNJ2013}
{Scargle}, J.~D., {Norris}, J.~P., {Jackson}, B., \& {Chiang}, J. 2013, \apj,
  764, 167, \dodoi{10.1088/0004-637X/764/2/167}

\bibitem[{{Spitler} {et~al.}(2016){Spitler}, {Scholz}, {Hessels}, {Bogdanov},
  {Brazier}, {Camilo}, {Chatterjee}, {Cordes}, {Crawford}, {Deneva}, {Ferdman},
  {Freire}, {Kaspi}, {Lazarus}, {Lynch}, {Madsen}, {McLaughlin}, {Patel},
  {Ransom}, {Seymour}, {Stairs}, {Stappers}, {van Leeuwen}, \&
  {Zhu}}]{Spitler2016}
{Spitler}, L.~G., {Scholz}, P., {Hessels}, J.~W.~T., {et~al.} 2016, \nat, 531,
  202, \dodoi{10.1038/nature17168}

\bibitem[{{Tsuzuki} {et~al.}(2024){Tsuzuki}, {Totani}, {Hu}, \&
  {Enoto}}]{TsuzukiTH2024}
{Tsuzuki}, Y., {Totani}, T., {Hu}, C.-P., \& {Enoto}, T. 2024, \mnras, 530,
  1885, \dodoi{10.1093/mnras/stae965}

\bibitem[{{Turolla} {et~al.}(2015){Turolla}, {Zane}, \&
  {Watts}}]{TurollaZW2015}
{Turolla}, R., {Zane}, S., \& {Watts}, A.~L. 2015, RPPh, 78, 116901,
  \dodoi{10.1088/0034-4885/78/11/116901}

\bibitem[{{van der Horst} {et~al.}(2012){van der Horst}, {Kouveliotou},
  {Gorgone}, {Kaneko}, {Baring}, {Guiriec}, {G{\"o}{\v{g}}{\"u}{\textcommabelow
  s}}, {Granot}, {Watts}, {Lin}, {Bhat}, {Bissaldi}, {Chaplin}, {Finger},
  {Gehrels}, {Gibby}, {Giles}, {Goldstein}, {Gruber}, {Harding}, {Kaper}, {von
  Kienlin}, {van der Klis}, {McBreen}, {Mcenery}, {Meegan}, {Paciesas},
  {Pe'er}, {Preece}, {Ramirez-Ruiz}, {Rau}, {Wachter}, {Wilson-Hodge}, {Woods},
  \& {Wijers}}]{vanderHorst2012}
{van der Horst}, A.~J., {Kouveliotou}, C., {Gorgone}, N.~M., {et~al.} 2012,
  \apj, 749, 122, \dodoi{10.1088/0004-637X/749/2/122}

\bibitem[{{van Putten} {et~al.}(2016){van Putten}, {Watts}, {Baring}, \&
  {Wijers}}]{2016MNRAS.461..877V}
{van Putten}, T., {Watts}, A.~L., {Baring}, M.~G., \& {Wijers}, R.~A.~M.~J.
  2016, \mnras, 461, 877, \dodoi{10.1093/mnras/stw1279}

\bibitem[{{Wadiasingh} {et~al.}(2018){Wadiasingh}, {Baring}, {Gonthier}, \&
  {Harding}}]{Wadiasingh2018}
{Wadiasingh}, Z., {Baring}, M.~G., {Gonthier}, P.~L., \& {Harding}, A.~K. 2018,
  \apj, 854, 98, \dodoi{10.3847/1538-4357/aaa460}

\bibitem[{{Wadiasingh} {et~al.}(2020){Wadiasingh}, {Beniamini}, {Timokhin},
  {Baring}, {van der Horst}, {Harding}, \& {Kazanas}}]{Wadiasingh2020}
{Wadiasingh}, Z., {Beniamini}, P., {Timokhin}, A., {et~al.} 2020, \apj, 891,
  82, \dodoi{10.3847/1538-4357/ab6d69}

\bibitem[{{Wadiasingh} \& {Timokhin}(2019)}]{Wadiasingh2019}
{Wadiasingh}, Z., \& {Timokhin}, A. 2019, \apj, 879, 4,
  \dodoi{10.3847/1538-4357/ab2240}

\bibitem[{{Wilms} {et~al.}(2000){Wilms}, {Allen}, \& {McCray}}]{WilmsAM2000}
{Wilms}, J., {Allen}, A., \& {McCray}, R. 2000, \apj, 542, 914,
  \dodoi{10.1086/317016}

\bibitem[{{Younes} {et~al.}(2015){Younes}, {Gogus}, {Kouveliotou}, \& {van der
  Hors}}]{Younes2015}
{Younes}, G., {Gogus}, E., {Kouveliotou}, C., \& {van der Hors}, A.~J. 2015,
  The Astronomer's Telegram, 7213, 1

\bibitem[{{Younes} {et~al.}(2014){Younes}, {Kouveliotou}, {van der Horst},
  {Baring}, {Granot}, {Watts}, {Bhat}, {Collazzi}, {Gehrels}, {Gorgone},
  {G{\"o}{\u{g}}{\"u}{\c{s}}}, {Gruber}, {Grunblatt}, {Huppenkothen}, {Kaneko},
  {von Kienlin}, {van der Klis}, {Lin}, {Mcenery}, {van Putten}, \&
  {Wijers}}]{YounesKv2014}
{Younes}, G., {Kouveliotou}, C., {van der Horst}, A.~J., {et~al.} 2014, \apj,
  785, 52, \dodoi{10.1088/0004-637X/785/1/52}

\bibitem[{{Younes} {et~al.}(2017){Younes}, {Kouveliotou}, {Jaodand}, {Baring},
  {van der Horst}, {Harding}, {Hessels}, {Gehrels}, {Gill}, {Huppenkothen},
  {Granot}, {G{\"o}{\u{g}}{\"u}{\textcommabelow s}}, \& {Lin}}]{YounesKJ2017}
{Younes}, G., {Kouveliotou}, C., {Jaodand}, A., {et~al.} 2017, \apj, 847, 85,
  \dodoi{10.3847/1538-4357/aa899a}

\bibitem[{{Younes} {et~al.}(2020){Younes}, {G{\"u}ver}, {Kouveliotou},
  {Baring}, {Hu}, {Wadiasingh}, {Begi{\c{c}}arslan}, {Enoto},
  {G{\"o}{\u{g}}{\"u}{\c{s}}}, {Lin}, {Harding}, {van der Horst}, {Majid},
  {Guillot}, \& {Malacaria}}]{Younes2020}
{Younes}, G., {G{\"u}ver}, T., {Kouveliotou}, C., {et~al.} 2020, The
  Astrophysical Journal Letters, 904, L21, \dodoi{10.3847/2041-8213/abc94c}

\bibitem[{{Younes} {et~al.}(2021){Younes}, {Baring}, {Kouveliotou},
  {Arzoumanian}, {Enoto}, {Doty}, {Gendreau}, {G{\"o}{\v{g}}{\"u}{\c{s}}},
  {Guillot}, {G{\"u}ver}, {Harding}, {Ho}, {van der Horst}, {Hu}, {Jaisawal},
  {Kaneko}, {LaMarr}, {Lin}, {Majid}, {Okajima}, {Pope}, {Ray}, {Roberts},
  {Saylor}, {Steiner}, \& {Wadiasingh}}]{YounesBK2021}
{Younes}, G., {Baring}, M.~G., {Kouveliotou}, C., {et~al.} 2021, Nature
  Astronomy, 5, 408, \dodoi{10.1038/s41550-020-01292-x}

\bibitem[{{Younes} {et~al.}(2022){Younes}, {Hu}, {Bansal}, {Ray}, {Pearlman},
  {Kirsten}, {Wadiasingh}, {G{\"o}{\u{g}}{\"u}{\c{s}}}, {Baring}, {Enoto},
  {Arzoumanian}, {Gendreau}, {Kouveliotou}, {G{\"u}ver}, {Harding}, {Majid},
  {Blumer}, {Hessels}, {Gawro{\'n}ski}, {Bezrukovs}, \&
  {Orbidans}}]{YounesHB2022}
{Younes}, G., {Hu}, C.-P., {Bansal}, K., {et~al.} 2022, \apj, 924, 136,
  \dodoi{10.3847/1538-4357/ac3756}

\bibitem[{{Younes} {et~al.}(2023){Younes}, {Baring}, {Harding}, {Enoto},
  {Wadiasingh}, {Pearlman}, {Ho}, {Guillot}, {Arzoumanian}, {Borghese},
  {Gendreau}, {G{\"o}{\v{g}}{\"u}{\c{s}}}, {G{\"u}ver}, {van der Horst}, {Hu},
  {Jaisawal}, {Kouveliotou}, {Lin}, \& {Majid}}]{Younes2023}
{Younes}, G., {Baring}, M.~G., {Harding}, A.~K., {et~al.} 2023, Nature
  Astronomy, 7, 339, \dodoi{10.1038/s41550-022-01865-y}

\bibitem[{{Zhang}(2023)}]{Zhang2023}
{Zhang}, B. 2023, Reviews of Modern Physics, 95, 035005,
  \dodoi{10.1103/RevModPhys.95.035005}

\bibitem[{{Zhang} {et~al.}(2020){Zhang}, {Xiong}, {Li}, {Li}, {Tuo}, {Ge},
  {Zhao}, {Xiao}, {Jia}, {Nie}, {Zhao}, {Luo}, {Li}, {Cai}, {Tan}, {Xue}, {Lu},
  {Song}, {Liu}, {Chen}, {Cao}, {Xu}, {Li}, {Lin}, \& {Zhang}}]{Zhang2020}
{Zhang}, S.~N., {Xiong}, S.~L., {Li}, C.~K., {et~al.} 2020, The Astronomer's
  Telegram, 13696, 1

\bibitem[{{Zhu} {et~al.}(2023){Zhu}, {Xu}, {Zhou}, {Lin}, {Wang}, {Wang},
  {Zhang}, {Niu}, {Chen}, {Li}, {Meng}, {Lee}, {Zhang}, {Feng}, {Ge},
  {G{\"o}{\u{g}}{\"u}{\c{s}}}, {Guan}, {Han}, {Jiang}, {Jiang}, {Kouveliotou},
  {Li}, {Miao}, {Miao}, {Men}, {Niu}, {Wang}, {Wang}, {Xu}, {Xu}, {Xue},
  {Yang}, {Yu}, {Yuan}, {Yue}, {Zhang}, \& {Zhang}}]{ZhuXZ2023}
{Zhu}, W., {Xu}, H., {Zhou}, D., {et~al.} 2023, Science Advances, 9, eadf6198,
  \dodoi{10.1126/sciadv.adf6198}

\end{thebibliography}

\appendix

\section{X-ray Burst Candidates Detected with \emph{NuSTAR}}\label{sec:appendix1}
We list start time, end time, peak time, T90, fluence, and hardness ratio in the appendix.

\begin{deluxetable*}{cccccccc}[h]

\tablecaption{Parameters of burst candidates detected with \nustar. Time system used in this table is barycentric corrected \nustar\ mission elapsed time.}
\tablehead{\colhead{$T_{\mathrm{start}}$} & \colhead {$T_{\mathrm{end}}$} & \colhead{$T_{\mathrm{peak}}$} & \colhead{$T_{\mathrm{90}}$} & \colhead {Fluence} & \colhead{Fluence Err} & \colhead {Hardness} & \colhead{Hardness Err}\\
\colhead{(s)} & \colhead{(s)} & \colhead{(s)} & \colhead{(s)} & \colhead{(erg~cm$^{-2}$)} & \colhead{(erg~cm$^{-2}$)} & \colhead{} & \colhead{}  }
\startdata
403409905.599507 & 403409906.960375 & 403409906.491022 & 1.3609 & 1.375E-09 & 3.815E-10 & 0.627 & 0.356 \\ 
403410000.364839 & 403410000.470499 & 403410000.396320 & 0.1057 & 1.175E-09 & 3.564E-10 & 0.570 & 0.354 \\ 
403410099.171249 & 403410100.301210 & 403410099.371248 & 1.1300 & 5.072E-08 & 3.158E-09 & 0.567 & 0.074 \\ 
403410222.543739 & 403410222.787176 & 403410222.644729 & 0.2434 & 8.333E-10 & 2.962E-10 & 3.024 & 2.475 \\ 
403410301.689052 & 403410302.129653 & 403410302.010395 & 0.4406 & 5.271E-10 & 2.359E-10 & 1.500 & 1.372 \\ 
$\ldots$ &  &  &  &  & &  &  \\ 
403453310.860278 & 403453312.677451 & 403491003.705208 & 0.1342 & 1.697E-09 & 4.267E-10 & 0.599 & 0.310 \\ 
403453325.201118 & 403453325.299989 & 403493457.540405 & 0.4100 & 5.191E-09 & 8.032E-10 & 0.618 & 0.197 \\ 
403464868.219885 & 403464871.119628 & 403497905.068596 & 0.4100 & 5.663E-09 & 8.183E-10 & 0.779 & 0.227 \\ 
403464908.423184 & 403464908.595032 & 403497907.078449 & 0.2364 & 1.712E-09 & 4.267E-10 & 0.449 & 0.241 \\ 
403464914.715458 & 403464916.915269 & 403502468.141533 & 0.0643 & 6.275E-10 & 2.560E-10 & 2.000 & 1.744 \\
\enddata
\tablecomments{This table is available in its entirety in machine-readable form.} 
\end{deluxetable*}

\end{document}